\documentclass[reprint,twocolumn,secnumarabic,amssymb, aps, prx]{revtex4-2}

\usepackage{graphicx}
\usepackage{gensymb}
\usepackage{amsmath, amssymb, amsthm}
\usepackage{braket}
\usepackage{lipsum}
\usepackage[colorlinks,linkcolor=black,citecolor=black,urlcolor=black,filecolor=black]{hyperref}
\usepackage{wrapfig}
\usepackage{listings}
\lstdefinestyle{mystyle}{
    basicstyle=\ttfamily\footnotesize,
    breakatwhitespace=false,         
    breaklines=true,                 
    captionpos=b,                    
    keepspaces=true,                 
    numbers=left,                    
    numbersep=5pt,                  
    showspaces=false,                
    showstringspaces=false,
    showtabs=false,                  
    tabsize=2
}
\usepackage[dvipsnames]{xcolor}
\usepackage{wasysym}
\usepackage{outlines}
\usepackage{enumerate}
\usepackage{setspace}
\usepackage{xcolor}
\usepackage[export]{adjustbox}
\usepackage{tikz}
\usepackage{multirow}
\usetikzlibrary{decorations.pathreplacing}
\usetikzlibrary{arrows,arrows.meta,calc,shapes.geometric,shapes.misc}
\tikzset{
	>=stealth',
	help lines/.style={dashed, thick},
	important line/.style={thick},
	connection/.style={thick, dotted},
}
\DeclareMathAlphabet{\mymathbb}{U}{BOONDOX-ds}{m}{n}
\usepackage{dsfont}
\usepackage{algorithm}
\usepackage{algpseudocode}
\usepackage[utf8]{inputenc}
\usepackage{verbatim}

\newcommand{\lpmemsmallparams}{\mbox{$[[2610, 744, \leq 16]]$}}
\newcommand{\lpmemmedparams}{\mbox{$[[4350, 1224, \leq 20]]$}}
\newcommand{\lpmemlargeparams}{\mbox{$[[5278, 1480, \leq 24]]$}}
\newcommand{\bbparams}{\mbox{$[[248, 10, \leq 18]]$}}
\newcommand{\lpprocparams}{\mbox{$[[1122, 148, \leq 20]]$}}
\newcommand{\bb}{\mbox{$\mathsf{bb}_{18}$}}
\newcommand{\lpmemlarge}{\mbox{$\mathsf{lp}_{24}^{3, 7}$}}
\newcommand{\lpmemmed}{\mbox{$\mathsf{lp}^{3, 7}_{20}$}}
\newcommand{\lpmemsmall}{\mbox{$\mathsf{lp}^{3, 7}_{16}$}}
\newcommand{\lpproc}{\mbox{$\mathsf{lp}^{3, 5}_{20}$}}

\newcommand{\nspacemed}{9,739}
\newcommand{\nbalancedmed}{11,961}

\newcommand{\nspacelarge}{11,033}
\newcommand{\nbalancedlarge}{13,255}

\newcommand{\ttoff}{\tau_\text{Toff.}}
\newcommand{\ttoffeccbalanced}{19\cdot\frac{2d}{3}}
\newcommand{\ttoffrsabalanced}{10\cdot\frac{2d}{3}}
\newcommand{\ttoffeccspace}{72\cdot\frac{2d}{3}}
\newcommand{\ttoffrsaspace}{43\cdot\frac{2d}{3}}

\newcommand{\trsatime}{97 days}

\newcommand{\teccbalanced}{264 days}
\newcommand{\tecctime}{10 days}

\newcommand{\ecc}{\mbox{ECC--256}}
\newcommand{\rsa}{\mbox{RSA--2048}}

\newcommand{\Nproc}{N_p}
\newcommand{\nproc}{n_p}
\newcommand{\kproc}{k_p}
\newcommand{\dproc}{d_p}
\newcommand{\Aproc}{\mathcal{A}_p}
\newcommand{\Nm}{N_m}
\newcommand{\nm}{n_m}
\newcommand{\km}{k_m}
\newcommand{\dm}{d_m}
\newcommand{\Am}{\mathcal{A}_m}
\newcommand{\Na}{N_{\text{\!}\mathcal{A}}}
\newcommand{\Nf}{N_f}
\newcommand{\nf}{n_f}
\newcommand{\kf}{k_f}
\newcommand{\df}{d_f}
\newcommand{\Af}{\mathcal{A}_f}
\newcommand{\Ns}{N_s}
\newcommand{\ns}{n_s}
\newcommand{\ks}{k_s}
\newcommand{\ds}{d_s}

\newcommand{\taus}{\tau_s}
\newcommand{\tauscult}{\tau_s^{\text{cult.}}}
\newcommand{\lket}[1]{\ket{\overline{#1}}}
\newcommand{\ccz}{\ket{\overline{\text{CCZ}}}}

\begin{document}

\title{Shor's algorithm is possible with as few as 10,000 reconfigurable atomic qubits}

\author{
Madelyn~Cain$^{1,*, \dagger}$, 
Qian~Xu$^{1, 2,*, \ddagger}$, 
Robbie King$^{1}$, 
Lewis~R.\,B.~Picard$^{1}$, 
Harry~Levine$^{1, 3}$, 
Manuel~Endres$^{1, 2}$, 
John~Preskill$^{1, 2}$, 
Hsin-Yuan~Huang$^{1, 2}$, 
Dolev~Bluvstein$^{1,2,\S}$
}

\affiliation{
\mbox{$^1$Oratomic, Pasadena, California 91125, USA}\\
\mbox{$^2$California Institute of Technology, Pasadena, California 91125, USA}\\
\mbox{$^3$Department of Physics, University of California, Berkeley, California 94720, USA}\\
\mbox{$^\dagger$mcain@oratomic.com, $^\ddagger$qxu@oratomic.com, $^\S$dbluvstein@oratomic.com}\\ 
\mbox{$^*$These authors contributed equally}}

\date{\today}

\begin{abstract}
Quantum computers have the potential to perform computational tasks beyond the reach of classical machines. A prominent example is Shor’s algorithm for integer factorization and discrete logarithms, which is of both fundamental importance and practical relevance to cryptography. However, due to the high overhead of quantum error correction, optimized resource estimates for cryptographically relevant instances of Shor’s algorithm require millions of physical qubits. Here, by leveraging advances in high-rate quantum error-correcting codes, efficient logical instruction sets, and circuit design, we show that Shor's algorithm can be executed at cryptographically relevant scales with as few as 10,000 reconfigurable atomic qubits. Increasing the number of physical qubits improves time efficiency by enabling greater parallelism; under plausible assumptions, the runtime for discrete logarithms on the P-256 elliptic curve could be just a few days for a system with 26,000 physical qubits, while the runtime for factoring \rsa\ integers is one to two orders of magnitude longer. Recent neutral-atom experiments have demonstrated universal fault-tolerant operations below the error-correction threshold, computation on arrays of hundreds of qubits, and trapping arrays with more than 6,000 highly coherent qubits. Although substantial engineering challenges remain, our theoretical analysis indicates that an appropriately designed neutral-atom architecture could support quantum computation at cryptographically relevant scales. More broadly, these results highlight the capability of neutral atoms for fault-tolerant quantum computing with wide-ranging scientific and technological applications. 
\end{abstract}

\maketitle

Quantum computers are believed to be capable of solving certain problems that are intractable for classical machines~\cite{deutsch1992rapid,simon1997power}. A prominent example is Shor’s algorithm, which provides superpolynomial speedups over known classical algorithms for integer factorization and discrete logarithms~\cite{shor1994algorithms,kitaev1995quantum}, problems that underpin widely used public-key cryptographic systems~\cite{rivest1978method, koblitz1987elliptic}. Discovered in 1994, this result provided the first evidence that quantum computers could outperform classical computation for problems of both fundamental and practical importance. 

The feasibility of large-scale quantum computation was initially questioned due to the fragility of quantum states, but this concern was addressed by pioneering work on quantum error correction in 1995 and 1996~\cite{shor1995scheme,calderbank1996good,shor1996fault,steane1996error,steane1996simple,gottesman1998theory,kitaev2003fault}, which showed that reliable quantum computation is possible provided physical error rates remain below a threshold value. These developments established the theoretical foundations for fault-tolerant quantum computers (FTQCs) capable of executing arbitrarily deep quantum circuits. Shor’s algorithm has since served as a flagship benchmark for large-scale quantum computation. Apart from applications to cryptanalysis, FTQCs are expected to enable wide-ranging applications across physics, materials science, quantum chemistry, machine learning, and beyond~\cite{dalzell2023quantum,huang2025vast,babbush2025grand,eisert2025mind}.

Despite these theoretical advances, realizing a utility-scale FTQC remains a major challenge. Current resource estimates for large-scale quantum computations---including cryptographically relevant instances of Shor’s algorithm---typically require on the order of millions of physical qubits~\cite{gidney2025how}. This scale is driven by the large overhead of quantum error correction, which often necessitates hundreds of physical qubits to encode a single logical qubit~\cite{fowler2012surface}. In contrast, today’s most advanced quantum processors are limited to at most a few hundred physical qubits. Closing this gap between algorithmic requirements and experimentally accessible hardware is therefore a central obstacle in quantum computing.

The past several decades have seen broad experimental efforts to develop FTQCs across a variety of physical platforms~\cite{postler2022demonstration, wang2023fault, acharya2024quantum, reichardt2024logical, ryan-anderson2024high, bluvstein2025architectural, reichardt2025fault}. Neutral-atom quantum computers~\cite{saffman2010quantum} in particular are rapidly emerging as a promising candidate for realizing a FTQC. The first digital quantum circuits were realized in 2022~\cite{bluvstein2022quantum, graham2022multi1}, leveraging parallel reconfigurability of atomic qubits, and shortly thereafter, error-corrected quantum algorithms were performed using up to hundreds of qubits~\cite{bluvstein2024logical, reichardt2024logical, bluvstein2025architectural, reichardt2025fault}. Despite this rapid progress, today's devices lag far behind the predicted requirements for practically useful quantum computers. Moreover, the comparatively slow clock speeds of neutral-atom systems are thought to potentially limit the practical implementation of deep circuits required for utility-scale processing. 

\begin{figure*}
    \centering
    \includegraphics[width=\linewidth]{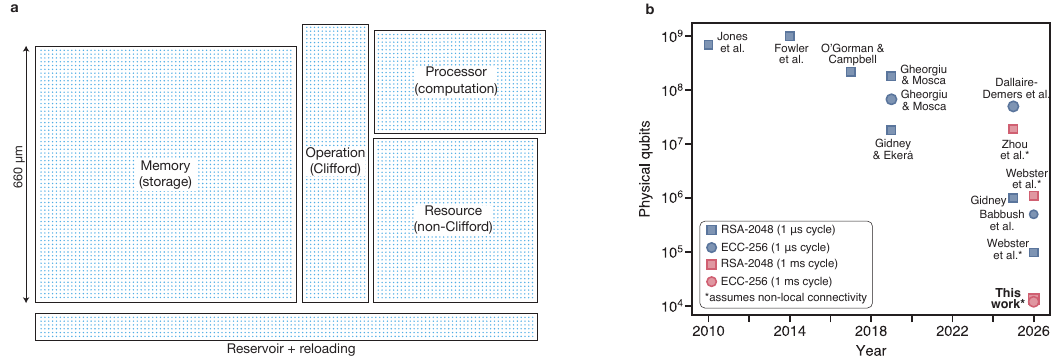}
    \caption{\textbf{Fault-tolerant computation with atomic qubits.} \textbf{a,} Neutral atom processor with four functional zones: a memory zone for storing quantum information, a processor zone for computation, an operation zone for performing Clifford operations, and a resource zone for generating magic states. Also included is a zone for an atomic reservoir and reloading. Each dot represents an atomic data qubit. \textbf{b,}~Estimated number of physical qubits to run Shor's algorithm versus year of publication for prior resource estimates~\cite{jones2012layered, fowler2012surface, ogorman2017quantum, gheorghiu2019benchmarking, gidney2019how, dallairedemers2025brace, gidney2025how, zhou2025resource, webster2026pinnacle} and the current work. 
    \label{fig:schematic}
    }
\end{figure*}

Here we propose and analyze architectures for FTQCs based on reconfigurable neutral-atom systems. Our scheme is based on high-rate error-correcting codes~\cite{Breuckmann2021, panteleev2021quantum, panteleev2022asymptotically, leverrier2022quantum, dinur2022good, bravyi2024high, xu2024constant}, which utilize nonlocal connectivity across a code block to protect many logical qubits with minimal overhead. By improving constructions for high-rate codes and decoders~\cite{hillmann2024localized}, we encode $\gtrsim$\,1{,}000 logical qubits with $\approx$\,30\% encoding rates at algorithmically relevant logical error rates. This enables qubit requirements to be reduced by an order of magnitude compared to architectures based on small, quasi-local codes with $\sim$\,4\% encoding rates~\cite{yoder2025tour, yang2026rascql, webster2026pinnacle}, and two orders of magnitude compared to planar surface-code architectures~\cite{gidney2025how}. Leveraging recent advances in logical operations with low space overhead~\cite{cohen2022low, cross2024improved, gidney2024magic, xu2025fast, zheng2025high}, we develop a verifiable compilation scheme and project further reductions in time overhead through parallel logical operations~\cite{zheng2025high, cuccaro2004new}. 

We explore a range of architectures with different numbers of physical qubits and parallelism, and evaluate them using prior state-of-the-art circuits for Shor’s algorithm~\cite{gidney2019how, chevignard2025reducing, gidney2025how, babbush2026quantum}, focusing on RSA with 2048-bit keys (\rsa) and elliptic-curve cryptography with 256-bit keys (\ecc). For elliptic-curve discrete logarithms, we explore architectures using approximately 10,000 to 26,000 physical qubits. Applying these architectures to recently developed low-depth circuits~\cite{babbush2026quantum}, we find runtimes varying by roughly two orders of magnitude, with the fastest constructions requiring as few as \tecctime, assuming a 1\,ms stabilizer measurement cycle time. For the RSA-2048 factoring problem, constructions using 11,000 to 14,000 qubits yield runtimes roughly two orders of magnitude longer due to  higher circuit depths~\cite{gidney2019how, gidney2025how}, while parallelized architectures with $\approx$\,102,000 qubits can potentially achieve runtimes of \trsatime.

Existing neutral atom systems have demonstrated below-threshold operation, universal fault-tolerant processing on up to 500 qubits~\cite{bluvstein2025architectural}, and trapping arrays exceeding 6,000 qubits~\cite{manetsch2025tweezer}. These advances are complemented by progress in continuous large-scale operation~\cite{gyger2024continuous,norcia2024iterative, chiu2025continuous,holman2026trapping} and high-fidelity operation across a variety of  atomic species~\cite{evered2023high,ma2023high,lis2023midcircuit, tsai2025benchmarking}. While substantial work is needed to integrate these advances into a complete apparatus and scale system sizes to the required levels, our analysis indicates that appropriately designed neutral-atom architectures could support cryptographically relevant implementations of Shor's algorithm. This finding underscores the importance of ongoing efforts to transition widely-deployed cryptographic systems to post-quantum standards designed to be secure against quantum attacks~\cite{FIPS203,FIPS204,FIPS205}. More broadly, we anticipate that neutral atom processors executing millions of gates on thousands of logical qubits will unlock a wide variety of applications with significant scientific and economic value.

\subsection*{Neutral-atom architecture\label{sec:architecture}}
Reconfigurable atom arrays offer unique advantages for quantum error correction because qubits can be dynamically rearranged during a computation. In this approach, physical qubits are encoded in long-lived clock states and stored in optical tweezer arrays generated by optical multiplexing devices~\cite{saffman2010quantum, Barredo2016, endres2016atom, ma2022universal,jenkins2022ytterbium, finkelstein2024universal}. High-fidelity entangling operations are performed via excitation to Rydberg states~\cite{isenhower2010demonstration, ma2023high, tsai2025benchmarking}. Between gate operations, the qubits are dynamically reconfigured by moving the optical tweezers, enabling massively parallel operation and nonlocal connectivity~\cite{Beugnon2007, Schlosser2011, bluvstein2022quantum}. Leveraging identical qubits and global parallel control simplifies the implementation of error correction protocols on redundant qubits~\cite{bluvstein2024logical}, and the parallel nonlocal connections enable both the use of transversal gates~\cite{bluvstein2024logical, cain2024correlated, zhou2024algorithmic, zhou2025resource} as well as high-rate encodings~\cite{bluvstein2024logical,xu2024constant, reichardt2025fault,zhang2025leveraging}. 

Figure~\ref{fig:schematic}a presents a conceptual schematic of an atomic quantum computer capable of implementing Shor's algorithm with \nbalancedmed\ qubits. The computer is divided into four primary functional zones, along with a zone for an atomic reservoir and reloading. The memory zone stores logical quantum information during the computation. The processor zone stores quantum information undergoing active computation. The operation zone is comprised of ancillary qubits used to perform Clifford logical  Pauli product measurements (PPMs), used for reading, writing, and editing of quantum information~\cite{cross2024improved, cowtan2025fast, li2025transversal, xu2025fast, zheng2025high}. Finally, the resource zone generates magic states to elevate Clifford PPMs to universal quantum computation. 

Feasible methods exist for operating atomic systems at the scales depicted in Fig.~\ref{fig:schematic}a and substantially beyond. Coherent arrays trapping up to 6,100 atomic qubits have already been demonstrated~\cite{manetsch2025tweezer} (without yet realizing quantum computation). Laser power overheads can be greatly reduced by holding atoms in optical potentials that are hundreds of times shallower except during imaging, cooling, and transport stages. Additionally, optical tweezer arrays with 360,000 traps have been demonstrated using metasurface phase masks (although without yet trapping atoms)~\cite{holman2026trapping}.

Furthermore, fault-tolerant architectures have been realized, incorporating simplified methods for removing all relevant entropy sources on systems of hundreds of qubits~\cite{bluvstein2025architectural}, and including continuous replenishment of lost qubits~\cite{gyger2024continuous,norcia2024iterative,chiu2025continuous}. Operation at 2$\times$ below threshold has also been demonstrated, with a path to 10$\times$ below threshold~\cite{bluvstein2025architectural}. Although high-fidelity entangling operations require high laser intensity and have thus far been realized only on regions of a few hundred qubits~\cite{evered2023high, ma2023high,lis2023midcircuit, tsai2025benchmarking, bluvstein2025architectural}, increasing laser power could directly enable operation on thousands of qubits. Further, the present experimental approach is limited by time-inefficient usage of laser power, where a continuous-wave laser is actively used with only a 0.1\% duty cycle ($\sim$\,200\,ns gate operations separated by idle periods of $\gtrsim$\,200\,$\mu$s). By rastering the beam dynamically to increase the active duty cycle, the number of atoms that can be addressed and entangled at high fidelity could thus be directly increased by three orders of magnitude. Although integrating these capabilities at larger scales requires substantial development effort, they appear to be mutually compatible, such that an appropriately designed architecture could realize the functionality illustrated in Fig.~\ref{fig:schematic}a.

Finally, demonstrated speeds with atomic processors are bottlenecked by speeds of readout and motion, which can be 100\,$\mu$s to several ms depending on the specific implementations~\cite{falconi2025microsecond}. Recent work, which did not optimize for speed, required several ms per stabilizer measurement cycle~\cite{bluvstein2025architectural}, albeit in a smaller system than considered in this work. Here we assume a 1\,ms stabilizer measurement cycle. Although realizing this in practice may require technological development, we anticipate that it can be achieved by optimizing for speed and leveraging the simple atomic motions inherited from the product structure of the codes we study~\cite{xu2024constant, bluvstein2024logical, bluvstein2025architectural}.

Figure~\ref{fig:schematic}b plots the resource estimates for cryptographically relevant problem instances as a function of time. Advances in logical codes, operations, and algorithms have reduced qubit requirements by five orders of magnitude over two decades. Note the differences between \rsa, the historical benchmark for Shor's algorithm, and \ecc. Elliptic curve cryptography is a more modern scheme which provides equivalent classical security, but with significantly smaller key sizes, resulting in reduced quantum complexity. Current state-of-the-art results for planar architectures estimate that \rsa\ requires  1 million qubits and 1 week~\cite{gidney2025factor}, whereas recent work finds that \ecc\ requires only half a million qubits and tens of minutes, assuming a 1\,$\mu$s cycle time~\cite{babbush2026quantum}. Here, we explore several architectures with differing resource requirements. We plot two such architectures for \rsa\ with \nbalancedlarge\ qubits, and \ecc\ with \nbalancedmed\ qubits (see the end of the resource estimates section for a detailed discussion of our results and prior work).

\begin{figure*}[ht!]
    \centering
    \includegraphics[width=\linewidth]{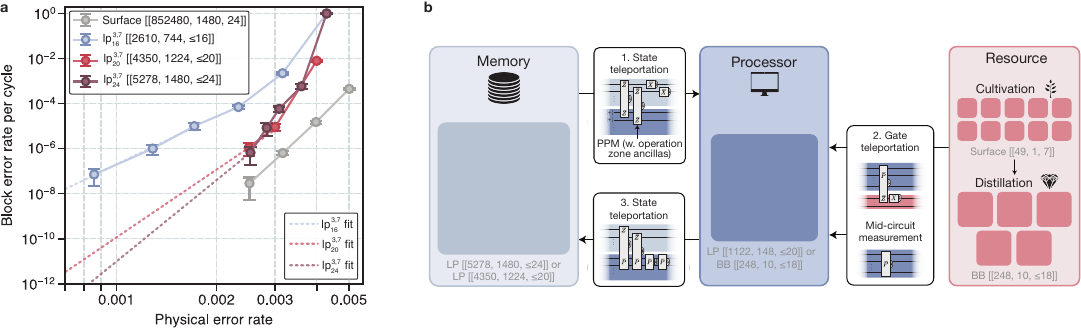}
    \caption{
    \textbf{Logical code performance and architecture.} \textbf{a,}~Block error rates per cycle for several lifted product codes and surface codes. Least-squares power law fits (dashed lines) are used to extrapolate to lower physical error rates $p$ which could not be numerically simulated. The blue fit is of the form $y=ax^b$, where $a=14.6\pm 0.7$ and $b=7.1 \pm 0.4$ are fitted parameters using data from the three smallest physical error rates. Using this same procedure, the fitted values of $b$ for \lpmemmed\ and \lpmemlarge\ are larger than $d/2$, the theoretical maximum value as $p\to 0$. To be conservative, we therefore fit the form $y=ax^{d/2}$ from the smallest physical error rate (red and purple). \textbf{b,}~Layout and compilation procedure for the logical architecture. The memory block stores quantum information, which is then teleported to the processor for computation. Sequential PPMs execute mid-circuit measurements and gate teleportation of magic states. Finally, the logical information is teleported back into the memory. Here, $\bar{P}$ denotes an arbitrary logical Pauli operator on the processor code.}
    \label{fig:codes}
\end{figure*}

\subsection*{Codes, logic, and compilation}
Reconfigurable atom arrays motivate architectures based on high-rate quantum low-density parity check (qLDPC) codes~\cite{breuckmann2021quantum}, which leverage nonlocality to densely pack many logical qubits into a single code block~\cite{bravyi2010tradeoffs}. Such codes can substantially reduce the overhead of error correction compared to the millions of qubits required for surface codes~\cite{xu2024constant, bravyi2024high, Breuckmann2021, panteleev2022asymptotically}. The main challenge is that, unlike the surface code setting, high-rate blocks make it harder to address individual logical qubits and to execute complex algorithms efficiently. Refs.~\cite{xu2024constant, bravyi2024high}, for example, provide concrete protocols for computation with high-rate qLDPC codes, but due to their recent nature, they were not yet highly optimized. Since then, key ingredients for architectures leveraging high-rate codes have improved substantially, particularly decoders~\cite{hillmann2024localized, muller2025improved} and logical operations~\cite{cross2024improved, williamson2024low, cowtan2025fast, he2025extractors, xu2025fast, malcolm2025computing, zheng2025high, li2025transversal, menon2025magic, xu2025batched}.

Figure~\ref{fig:codes}a plots the block error rates of high-rate codes we develop here, leveraging improved quasi-cyclic lifted-product (LP) code~\cite{panteleev2019degenerate} constructions with encoding rates of approximately $30\%$. We analyze three instances from this family with parameters $\lpmemsmallparams$, $\lpmemmedparams$, and $\lpmemlargeparams$, where $[[n, k, d]]$ denotes a code with $n$ physical qubits, $k$ logical qubits, and distance $d$. We denote these codes as \lpmemsmall, \lpmemmed, and \lpmemlarge, respectively, where the subscript and superscript correspond to the code distance and seed matrix dimensions respectively; see Appendix~\ref{app:codes}. We use a circuit-level noise model where entangling gates, state preparations, and measurements experience depolarizing noise with error rate $p$, and decode with a customized belief propagation and localized statistics decoder~\cite{hillmann2024localized}~(Appendix~\ref{app:numerics}). At \mbox{$p=0.1\%$}, the \lpmemlarge\ code achieves extrapolated per-cycle block failure rates (i.e. the probability that any logical qubit fails) of  $\approx$\,$10^{-11}$. Further, at low physical error rates, its block error rates are comparable to surface codes with the same distance and number of logical qubits, but with 161\,$\times$ fewer physical qubits. We note that a hardware-realistic error model with $>$\,50\% of errors due to heralded atom loss and biased Pauli noise could effectively lower $p$ by a factor of two~\cite{wu2022erasure,sahay2023high,kubica2023erasure,yu2024processing,chang2024surface,perrin2024quantum, baranes2025leveraging, gu2025fault, bluvstein2025architectural}, further reducing the block error rates. We also anticipate that the current performance is decoder-limited and that further error suppression can be achieved using more advanced decoders~\cite{muller2025improved, senior2026scalable, maan2026decoding, ataides2025neural}.

With these high-rate codes, one can perform simple resource estimates for a general algorithm. Clifford operations~\cite{bravyi2016trading} can in principle be implemented directly on a large qLDPC block (e.g., $\lpmemlarge$) using code-surgery techniques~\cite{cohen2022low, cross2024improved, he2025extractors}. These methods use an ancillary system to fault-tolerantly measure an arbitrary logical Pauli operator of a general qLDPC code. Magic $\lket{T}$ states may be generated, for example, using surface-code cultivation~\cite{gidney2024magic}, at an additional space cost of $\approx$\,500 qubits. The algorithm can then be executed using Pauli-based computation~\cite{bravyi2016trading}, in which Clifford gates are propagated to the end of the circuit, transforming Pauli-product measurements (PPMs) arising from $\overline{T}$-gate teleportation~\footnote{A $\overline{T}$ gate on the memory code can be implemented using a two-qubit $\bar{Z}\bar{Z}$ measurement between the memory and the surface code hosting the $\lket{T}$ state~\cite{litinski2019game}.} and mid-circuit Pauli measurements into higher-weight logical PPMs, potentially acting on all logical qubits in the code block. In this scheme, each PPM is sequentially implemented via a surgery gadget, such that the total runtime is proportional to the Toffoli count of the algorithm. Each Toffoli gate requires approximately $\ttoff \simeq 4d$ stabilizer measurement cycles, corresponding to 4–7 $\lket{T}$-state teleportations per Toffoli gate~\cite{jones2013low, matthew2013meet, gidney2018halving}.

Crucially, however, the complication in actually executing logic on high-rate codes hides inside of the code surgery implementation, which can present substantial challenges. Such fault-tolerant surgeries can, in principle, be implemented for any qLDPC code, with an ancilla overhead scaling with the block size~\cite{cross2024improved, he2025extractors}. However, their practical construction and optimization for large codes, while maintaining low qubit overhead and rigorously benchmarking performance, remain technically challenging and computationally demanding~\footnote{State-of-the-art, low-overhead constructions rely on randomized numerical subroutines that extensively estimate the fault distance of the protocol~\cite{cross2024improved, zheng2025high}, a procedure that becomes computationally prohibitive for large code blocks.}. For this reason, we study below an architecture based on verifiable code surgeries as a theoretically consistent existence proof, which may not be optimal in terms of resource costs. This is achieved by performing computation on smaller high-rate ``processor'' codes, rather than directly on the large codes in Fig.~\ref{fig:codes}a.

Figure~\ref{fig:codes}b shows the resulting architecture where various qLDPC codes are used for memory, processing, and resource-state generation. An additional operation zone contains ancillary qubits for performing code surgeries. We use either the \lpmemmed\ or the \lpmemlarge\ code for the memory, depending on the number of logical qubits required for a given algorithm. We consider two possible processor codes: a \bbparams\ bivariate bicycle code~\cite{liang2025generalized}, denoted \bb, and a \lpprocparams\ LP code, denoted \lpproc\ (Appendix~\ref{app:codes}). We refer to architectures with these two choices of the processor code as space-efficient and balanced, respectively, as the \bb\ code has reduced space cost and logical qubits (and therefore processing power) compared to the \lpproc\ code. Finally, because arithmetic circuits in cryptographic algorithms naturally use non-Clifford Toffoli gates, we generate $\ccz$ resource states to directly perform Toffoli gates. We use high-rate $8T$-to-CCZ distillation~\cite{litinski2019magic, xu2025fast} to distill cultivated $\lket{T}$ states~\cite{gidney2024magic, sahay2025fold} into many $\ccz$ states hosted in \bb\ factory codes in parallel. Each resource state has logical error rate $\lesssim$\,$10^{-10}$ at $p=0.1\%$, and is generated in time less than a single surgery cycle on average (Appendix~\ref{app:magic}). 

Universal computation can be performed on this architecture using the following compilation strategy, detailed in Appendix~\ref{app:compilation} and illustrated in Fig.~\ref{fig:codes}b. A circuit, comprised of Clifford and Toffoli gates, is first decomposed into sub-circuits $\{C_i\}$ acting on $m_i$ qubits, where each sub-circuit fits inside the processor block and contains $\beta_i$ Toffoli gates, $\gamma_i$ mid-circuit Pauli measurements, and arbitrary Clifford gates. Each $C_i$ is then executed sequentially as follows:
\begin{enumerate}
    \item \emph{Teleport to processor.} 
    The $m_i$ logical qubits are teleported from the memory code to the processor code using $2m_i$ PPMs: $m_i$ inter-zone measurements between memory and processor, followed by $m_i$ single-qubit measurements on the memory.
    \item \emph{Pauli-based computation on the processor.} 
    Clifford gates are absorbed into high-weight PPMs when teleporting in $\ccz$ states for Toffoli gates and when implementing mid-circuit measurements. 
    \item \emph{Teleport to memory.} The $m_i$ qubits are teleported back using another $2m_i$ PPMs: $m_i$ inter-zone measurements (now Clifford-transformed) followed by $m_i$ measurements on the processor.
\end{enumerate}

As described in Appendix~\ref{app:surgery}, these PPMs can be implemented using standard code surgery techniques~\cite{cross2024improved, williamson2024low, swaroop2024universal} by reconfiguring operation zone ancillary qubits to measure the chosen logical operators. Importantly, the ancilla qubit count scales with the maximum physical weight of the target logical operators. All PPMs in our scheme are of the form $\bar{Z}_i\bar{P}$ or $\bar{X}_j$, where $\bar{Z}_i$ (resp. $\bar{X}_j$) is a single-qubit logical $\bar{Z}$ (resp. $\bar{X}$) operator on the memory or factory code, and $\bar{P}$ is an arbitrary logical operator on the processor code (Fig.~\ref{fig:codes}b). Because the logical operators in the memory and factory codes can be chosen to have low physical weight~\cite{zheng2026high}, the ancilla cost therefore scales with the processor code size, rather than with the entire memory block, as would be required if complex operations were performed directly on the memory code. We numerically construct the required surgery gadgets for our architecture and estimate the corresponding ancilla costs, and hence the size of the operation zone, in Appendix~\ref{app:surgery}, finding it accounts for only $10\%$--$20\%$ of the total qubit count in both the space-efficient and balanced architectures. Numerical simulations indicate that the logical error rates of the surgery gadgets are within an order of magnitude of the processor and factory code, comparable to the \lpmemmed\ and \lpmemlarge\ error rates in Fig.~\ref{fig:codes}a. Therefore, we take the memory zone error rate as an approximation for the total architecture error rate.

The time cost of this compilation strategy is as follows. Each PPM is implemented in a single \emph{surgery cycle} consisting of $2d/3$ stabilizer measurement cycles, chosen to minimize the total logical error rate (see Ref.~\cite{cross2024improved} and Appendix~\ref{app:surgery} for justification), where $d$ is the processor code distance. Each  $C_i$ involves \mbox{$4m_i + 4\beta_i + \gamma_i$} PPMs, thus taking \mbox{$\frac{2d}{3}(4m_i + 4\beta_i + \gamma_i)$} cycles. As the runtime of this compilation strategy scales with the total Toffoli count, we compute the amortized time per Toffoli $\ttoff$ of the circuit. $\ttoff$ is minimized when each $C_i$ contains many Toffoli gates relative to mid-circuit measurements and communication between the memory and processor (\mbox{$\beta_i \gtrsim m_i, \gamma_i$}), i.e. \mbox{$\ttoff=\frac{2d}{3\beta_i}(4m_i + 4\beta_i + \gamma_i) \geq \frac{8d}{3}$.} In practice, however, the exact cost depends on the concrete logical circuit. For Shor's algorithm we therefore estimate $\ttoff$ based on the ripple-carry adder~\cite{vedral1996quantum, cuccaro2004new, gidney2018halving} and unary lookup table~\cite{babbush2018encoding}, which we anticipate dominate the Toffoli counts in the circuits we consider~\cite{gidney2019how, chevignard2025reducing, gidney2025how, proos2003shor, roetteler2017quantum, haner2020improved, litinski2023compute, gouzien2023performance, babbush2026quantum}. As shown in Appendix~\ref{app:compilation}, for the balanced architecture we estimate \mbox{$\ttoff \approx\ttoffeccbalanced$} and \mbox{$\ttoff\approx\ttoffrsabalanced$} cycles for \ecc~\cite{babbush2026quantum} and \rsa~\cite{gidney2019how, gidney2025how}, respectively. For the space-efficient architecture, the reduced processor size increases the time to \mbox{$\ttoff \approx \ttoffeccspace$} and \mbox{$\ttoff \approx \ttoffrsaspace$} cycles for \ecc~\cite{babbush2026quantum} and \rsa~\cite{gidney2019how, gidney2025how}, respectively. 

The space cost is computed by summing the number of qubits in each zone. We count the data qubits as well as one basis ($X$ or $Z$) of ancilla qubits for measuring stabilizers by assuming that the $X$- and $Z$-stabilizers are measured sequentially. Using the \lpmemmed\ memory code, the resulting qubit counts are \nspacemed\ and \nbalancedmed\ for the space-efficient and balanced architectures, respectively, whereas for \lpmemlarge\ memory code, the respective qubit counts are \nspacelarge\ and \nbalancedlarge.

\begin{figure*}
    \centering
    \includegraphics{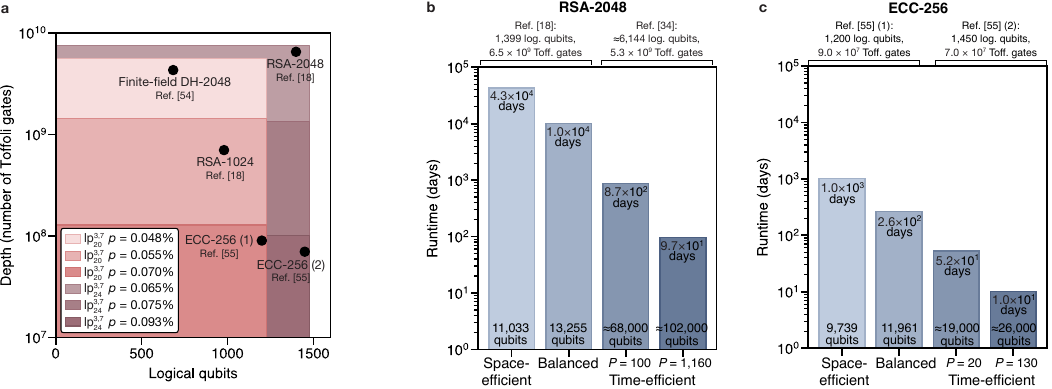}
    \caption{\textbf{Resources required for Shor's algorithm.} \textbf{a,}~Number of Toffoli gates and logical qubits required in prior circuits for \mbox{RSA--1024} and \rsa~\cite{gidney2025how}, finite-field \mbox{DH--2048}~\cite{chevignard2025reducing}, and \ecc~\cite{babbush2026quantum}. The colored boxes represent the total number of Toffoli gates that can be executed in the balanced architecture  with $90\%$ total success probability, using either the \lpmemmed\ or \lpmemlarge\ memory. \textbf{b,}~Runtimes required for \rsa\ in different architectures (annotations have one significant figure). The space-efficient and balanced architectures use the \lpmemlarge\ memory and  the circuit in Ref.~\cite{gidney2025how}, whereas the time-efficient architecture uses the circuit in Ref.~\cite{gidney2019how}. We emphasize that the time-efficient runtimes are based on the assumption that up to $P$ Toffoli gates are executed in parallel using high-rate surgery, as motivated by Ref.~\cite{zheng2025high}. \textbf{c,}~Runtimes required for solving \ecc, using the compilations in Ref.~\cite{babbush2026quantum}. The space-efficient and balanced architectures use the \lpmemmed\ memory. 
    }
    \label{fig:resources}
\end{figure*}

\subsection*{Resource estimates}
We now determine the resources required to run Shor's algorithm, focusing on practically relevant benchmarks including Diffie-Hellman (DH), RSA, and ECC. Figure \ref{fig:resources}a illustrates the required number of logical qubits and Toffoli gate count for prior state-of-the-art circuits~\cite{chevignard2025reducing, gidney2025how, babbush2026quantum}. We include both compilations for \ecc\ from Ref.~\cite{babbush2026quantum}, denoted (1) and (2), which offer a tradeoff between logical qubit and Toffoli count. Also plotted are the number of Toffoli gates which can be implemented with 90\% total success probability in the balanced architecture, \mbox{$n_\text{Toff.} = \log(0.9)/[\ttoff\log(1-P_L)]$}, where \mbox{$\ttoff = \ttoffeccbalanced$}, and $P_L$ is the extrapolated block error rate per cycle of the memory code. We consider both the \lpmemmed\ and \lpmemlarge\ memory blocks, which can implement circuits with \mbox{$\leq1224$} and \mbox{$\leq1480$} logical qubits, respectively. \ecc\ requires \mbox{$p=0.093\%$} with the \lpmemlarge\ memory (compilation~(1) can also be implemented with \mbox{$p=0.070\%$} using the smaller \lpmemmed\ memory), whereas \rsa\ and DH--2048 require slightly reduced physical error rates.

In Figure~\ref{fig:resources}b-c, we plot the number of days required for \rsa\ and \ecc, assuming 1\,ms per cycle. The time costs vary substantially depending on the logical architectures, circuits, and algorithms. For example, the runtimes of \rsa~(Fig.~\ref{fig:resources}b) and \ecc~(Fig.~\ref{fig:resources}c) vary by nearly two orders of magnitude for the balanced architecture, with the latter requiring $\approx$\,\teccbalanced. The runtimes for both the space-efficient and balanced architectures are limited by the fact that Toffoli gates and PPMs are executed sequentially. By changing the fault-tolerant schemes for implementing logic and compiling circuits to leverage parallelism, further runtime reductions by nearly two orders of magnitude can potentially be achieved with the time-efficient architecture plotted.

Concretely, in the time-efficient architecture, we consider parallelizing the Toffoli gates in core algorithmic subroutines assuming the use of parallel surgery operations~\cite{zhang2025time, cowtan2025parallel, zheng2025high}. Such surgery operations have previously been constructed for hundreds of parallel PPMs on related code families, and have demonstrated comparable performance to an idling code block in numerical benchmarks on $\approx$\,10 PPMs~\cite{zheng2025high}. We estimate the potential speedup by computing the time reduction from replacing $n$-bit serial quantum ripple-carry adders~\cite{vedral1996quantum, cuccaro2004new, gidney2018halving}, commonly used in prior circuits~\cite{gidney2019how, gidney2025how}, with parallelized carry-lookahead adders~\cite{draper2004logarithmic} (here $n$ is the key size). These adders have Toffoli depths (number of maximally parallelized layers of Toffoli gates) of \mbox{$\approx$\,$1n$--$2n$} and $\approx$\,$4\log(n)$, respectively, where the range corresponds to whether a particular addition is controlled or not. We consider the \ecc\ and \rsa\ compilations in Refs.~\cite{gidney2019how, babbush2026quantum}. As detailed in Appendix~\ref{app:time_efficient}, we generate and consume $\ccz$ states in parallel batches of size $P$, where the gate teleportation and $\overline{\text{CZ}}$ feedback are directly implemented in three parallel surgery cycles~\cite{horsman2012surface, zheng2025high}. The estimated speedup increases with $P$, with projected runtimes of \tecctime\ for \ecc\ ($P=130$) and \trsatime\ for \rsa\ ($P=1{,}160$).

To realize these time improvements, the number of physical qubits must be increased to generate more $\ccz$ states in parallel, and for ancillary logical qubits used in the carry-lookahead adder~\cite{draper2004logarithmic}. We use codes in upcoming work~\cite{caltech2026codeatlas}, which have 20\%--30\% encoding rates and maintain the required distance, for storing quantum information, computation, and generating high-rate magic. By assuming that the surgery system scales with the size of codes undergoing parallel operation, in line with prior work~\cite{zheng2025high}, we estimate that approximately 26,000 qubits are required for \ecc\ (\mbox{$P=130$}) and 102,000 qubits are required for \rsa\ (\mbox{$P=1{,}160$}) (see Appendix~\ref{app:time_efficient} for  discussion). We emphasize that these estimates are preliminary and can be  optimized in future work. Nonetheless, they indicate that the runtime can be substantially reduced by optimizing the algorithm, circuits, and fault-tolerant protocols.

Finally, the advances in this work complement recent progress in the resource estimates plotted in Fig.~\ref{fig:schematic}b. Advances in planar surface-code architectures and algorithmic optimizations have already reduced qubit requirements to under one million qubits~\cite{gidney2025factor, babbush2026quantum}. Recent progress across hardware platforms has also motivated architectures leveraging nonlocal connectivity to reduce resource costs~\cite{zhou2025resource, yoder2025tour, webster2026pinnacle}. Zhou \textit{et al.}~\cite{zhou2025resource} compile \rsa\ with 19 million qubits encoded in surface codes, and uses fast, nonlocal transversal operations to achieve a runtime of $\approx$\,1 week despite a slower 1\,ms cycle time. Architectures based on high-rate codes encoding small $\approx$\,10 logical qubits have also been proposed, achieving 10$\times$ qubit savings over planar architectures while maintaining fixed, quasi-local connectivity and enabling parallel logical operations across blocks~\cite{yoder2025tour, webster2026pinnacle}. For example, Webster \textit{et al.}~\cite{webster2026pinnacle} compile Shor's algorithm for \rsa\ with various resource tradeoffs; we plot two such compilations with 1.1 million qubits and 1 year (1\,ms cycle time), and 98,000 qubits and 1 month (1\,$\mu$s cycle time). Our space-efficient and balanced architectures further reduce qubit counts by one order of magnitude compared to small-block architectures by leveraging the long-range, reconfigurable connectivity of neutral-atom systems. While we expect these architectures have near-optimal qubit counts, our findings indicate that runtimes can be reduced by orders of magnitude by leveraging parallelism~\cite{draper2004logarithmic, zhang2025time}, suggesting that future efforts should focus on architectural design and reducing algorithm runtime.

\subsection*{Conclusion and outlook}

In this work, we propose FTQC architectures based on reconfigurable atom arrays. By optimizing constructions for high-rate codes~\cite{Breuckmann2021, panteleev2021quantum, panteleev2022asymptotically, leverrier2022quantum, dinur2022good, bravyi2024high, xu2024constant}, decoders~\cite{hillmann2024localized}, and low-overhead logical operations~\cite{cohen2022low, cross2024improved, gidney2024magic, xu2025fast, zheng2025high}, we find that Shor's algorithm can be implemented at cryptographically relevant scales using as few as 10,000 atomic qubits. Leveraging improvements to quantum algorithms~\cite{babbush2026quantum} and parallel logical operations and circuits~\cite{zhang2025time, draper2004logarithmic}, our most time-efficient architectures can potentially enable runtimes of  \tecctime\ for \ecc\ with $\approx$\,26,000 qubits, and \trsatime\ for \rsa\ with $\approx$\,102,000 qubits. 

The space and time overheads for these problems are likely to further improve. The space overheads could potentially be reduced by a factor of two beyond our most space-efficient architectures by utilizing higher-rate codes~\cite{caltech2026codeatlas}. Substantially larger space reductions, however, would likely require algorithmic compression and improved physical fidelities that permit operation at lower code distances. In contrast, advances in quantum error correction, circuits, and algorithms could potentially reduce runtimes by more than an order of magnitude beyond the time-efficient architecture presented. Each parallel Toffoli layer can potentially be implemented in $\simeq$\,1 cycle by using single-shot surgery~\cite{baspin2025fast, cowtan2025fast, xu2025batched, chang2026constant}, single-shot code switching~\cite{tan2025single, li2025transversal, golowich2025constant}, or transversal gates~\cite{cain2024correlated, zhou2024algorithmic, bombin2013gauge, brown2016fault, menon2025magic, jacob2026single, xu2025fast}, which could remove the requirement for \mbox{$d\simeq 20$} cycles per operation. Computation can potentially be further parallelized using space-time tradeoffs~\cite{fowler2012bridge, fowler2012time, xu2024fast}, advanced compilation strategies~\cite{litinski2022active, webster2026pinnacle}, and improvements to high-rate magic~\cite{menon2025magic, li2025transversal}. While we emphasize that such schemes require further innovation and development, they open the possibility of reducing runtime by at least another order of magnitude. 

The hardware itself can also be improved along several fronts to reduce runtime. Although systems of $\approx$\,10,000 qubits are sufficient to address the relevant problem sizes, known techniques could potentially increase the available space to $\sim$\,100,000 qubits. Space-time tradeoffs can then be leveraged to reduce computation time by factors of up to 6--10\,$\times$~\cite{fowler2012bridge, fowler2012time}. Raw qubit speeds can also be improved substantially. Faster readout techniques can reduce measurement times from $\sim$1\,ms to $\sim$1\,$\mu$s~\cite{falconi2025microsecond}. Faster qubit motion can be achieved through architectures that leverage atoms moving at \textit{constant velocity}, thereby avoiding repeated starting and stopping and enabling transport speeds that are potentially hundreds of times faster. These approaches require additional work to incorporate into a complete architecture that removes all forms of entropy~\cite{bluvstein2025architectural}. Nevertheless, they indicate that hardware-level speed improvements by several additional orders of magnitude can potentially be achieved, resulting in runtimes on the scale of hours or even minutes.

These findings have significant implications. Although substantial expertise, experimental development effort, and architectural design are required, our theoretical analysis suggests that a neutral atom system capable of implementing Shor's algorithm could be constructed. This conclusion underscores the importance of ongoing efforts to transition widely-deployed cryptographic systems toward post-quantum standards designed to be secure against quantum attacks~\cite{FIPS203,FIPS204,FIPS205}. More broadly, these results position neutral-atom systems as a leading platform for utility-scale quantum computation, with the capacity to drive innovation across science and industry. \\

\clearpage
\newpage

\bibliographystyle{naturemag_arxiv2.bst}
\bibliography{main.bbl}
\clearpage
\newpage

\setcounter{figure}{0}
\newcounter{EDfig}
\renewcommand{\figurename}{Extended Data Fig.}
\renewcommand{\tablename}{Extended Data Table}

\section*{Methods}
\appendix
Here we provide technical details on the codes, logic, and compilation schemes discussed in this work. In Appendix~\ref{app:codes}, we describe and numerically benchmark the codes used in this work, including the lifted product code constructions and bivariate bicycle code constructions. In Appendix~\ref{app:surgery}, we describe how code surgery is performed, including a general description, the concrete constructions, and the associated resource costs and opportunities for future improvement. Appendix~\ref{app:magic} describes the high-rate distillation procedure and the associated resource costs. Appendix~\ref{app:numerics} includes details of the numerical simulations. In Appendix~\ref{app:compilation}, we describe the compilation procedure for the space-efficient and balanced architectures, and their application to ripple-carry adders~\cite{cuccaro2004new, gidney2018halving} and table lookup~\cite{babbush2018encoding} circuits. Finally, Appendix~\ref{app:time_efficient} describes the time-efficient architecture and the associated resource costs. Extended Data Table~\ref{tab:notation} summarizes the notation for different quantities used in this work.

\begin{table*}[ht!]
\caption{\textbf{Notation summary.} Summary of notation used in this work. \label{tab:notation}}
\begin{ruledtabular}
\begin{tabular}{cp{13.5cm}} 
 Notation & Description \\ \hline
    $[[\nm, \km, \dm]]$ & Memory code parameters for the space-efficient and balanced architectures. \\
    $[[\nproc, \kproc, \dproc]]$ & Processor code parameters for the space-efficient and balanced architectures. \\
    $[[\nf, \kf, \df]]$ & Factory code parameters  used in high-rate 8$T$-to-CCZ distillation. \\
    $[[\ns, \ks, \ds]]$ & Surface code parameters  used in magic state cultivation. \\
  $N_i$ & Size of a code indexed by $i$, including the data qubits and one basis ($X$ or $Z$) of stabilizers, approximated by \mbox{$N_i\simeq n_i + \lfloor (n_i-k_i)/2\rfloor$}. \\
  $\Nm$ & Size of the memory code. \\
  $\Nproc$ & Size of the processor code. \\
  $\Nf$ & Size of the factory codes. \\
  $\Ns$ & Size of the cultivated surface codes. \\
  $\Na$ & Size of the operation zone ancilla system. \\
  $\Am$ & Ancilla system for the memory code. \\
  $\Aproc$ & Ancilla system for the processor code. \\
  $\Af$ & Ancilla system for the factory codes. \\
  $\taus$ & Number of stabilizer measurement cycles during surgery PPMs, equal to $\frac{2}{3}\dproc$. \\
  $\tauscult$ & Number of stabilizer measurement cycles to teleport cultivated  states during high-rate distillation, equal to $\frac{2}{3}\ds$. \\
  \lpmemsmall & Small LP memory code with parameters \lpmemsmallparams. \\
   \lpmemmed & Medium LP memory code with parameters \lpmemmedparams. \\
   \lpmemlarge & Large LP memory code with parameters \lpmemlargeparams. \\
   \lpproc & LP processor code for the balanced architecture with parameters \lpprocparams. \\
   \bb & BB factory code and processor code for the space-efficient architecture with parameters \bbparams. 
\end{tabular}
\end{ruledtabular}
\end{table*}

\section{Codes} \label{app:codes}
Here we describe the high-rate quantum error-correcting codes used in this work and their functional roles. The codes are based on optimized constructions of lifted-product (LP) codes~\cite{panteleev2019degenerate, xu2024constant} and prior constructions of bivariate bicycle (BB) codes~\cite{panteleev2021quantum, symons2025sequences}. As summarized in Extended Data Table~\ref{tab:all_codes}, these codes offer different tradeoffs between block size, encoding rate, distance, and stabilizer weight. Here, the weight of a Pauli operator refers to the number of qubits for which it has nonzero support. These parameters are relevant to their practical performance, as the code distance bounds the logical error scaling with decreasing physical error rates~\cite{fowler2012surface}, and the stabilizer weight is related to how many entangling gates are needed for stabilizer measurements.

\subsection{Lifted-product codes\label{app:lp_codes}}

The LP code family~\cite{panteleev2019degenerate} is obtained by taking the product of two classical codes with check matrices \mbox{$A \in R^{r_A \times n_A}$} and $B \in R^{r_B \times n_B}$, where $R := \mathbb{F}_2[x]/(x^\ell + 1)$ is the univariate polynomial ring of order $\ell$. Each entry of $A$ is an element of $R$, which can be identified with a polynomial of degree at most $\ell - 1$ over $\mathbb{F}_2$, or equivalently, with an $\ell \times \ell$ circulant permutation matrix. In this work, we always choose $B = A^{\dagger}$, where $A^{\dagger}$ denotes the transpose of $A$ with each polynomial entry $p(x)$ replaced by \mbox{$p(x^{-1}) \mod (x^\ell + 1)$}. As the circulant permutation matrix corresponds to a cyclic shift in a line of atoms, this family of codes is readily implementable on reconfigurable atom arrays; see Ref.~\cite{xu2024constant} for their detailed description and proposed implementation.

The resulting quantum code $\mathrm{LP}(A, A^{\dagger})$ has parameters
\begin{equation}
  [[n = (r_A^2 + n_A^2)\,\ell,\quad k \geq (n_A - r_A)^2\,\ell,\quad d\,]], \label{eq:lp_parameters}
\end{equation}
with stabilizer weight $r_A + n_A$, where the distance satisfies the inequality~\cite{smarandache2012quasi, raveendran2025minimum}
\begin{equation}
    d \leq \min((r_A+1)!, \quad (n_A+1)!). \label{eq:lp_distance_bound}
\end{equation}
As notation, we refer to an LP code built from an \mbox{$r_A \times n_A$} seed matrix $A$ over a ring of order $\ell$ and achieving quantum distance $\leq d$ as $\mathsf{lp}^{r_A, n_A}_d$. 

In this work, we study codes with \mbox{$r_A\times n_A=3\times 7$} and $3\times5$ seed matrices, corresponding to a distance of $d\leq 24$ from Eq.~\eqref{eq:lp_distance_bound}. Because the true distance can be lower than this bound, we numerically obtain improved estimates of the distance upper bound. We use a decoder-based distance estimation~\cite{bravyi2024high} using the BP-OSD decoder \cite{roffe2020decoding} for both the distance of the $Z$-check matrix ($d_Z$) and the $X$-check matrix ($d_X$). Concretely, for $d_Z$ (and similarly for $d_X$), let $L_X \in \mathbb{F}_2^{k \times n}$ be a basis for the logical $X$ operators $\ker H_X / \mathrm{im}\, H_Z^\top$. For each trial, we sample $c \sim \{0,1\}^k \setminus \{0\}$, set $e = c \cdot L_X \pmod{2}$, and use BP-OSD to find a minimum-weight $v$ satisfying $H_X v = 0$ and $e \cdot v = 1 \pmod{2}$. Then $d_Z \leq \min_t |v_t|$. For this method, we consider at least $25,000$ trials for both $d_Z$ and $d_X$. Below we describe the LP code instances studied in this work. The \lpproc\ processor code and the memory codes \lpmemsmall, \lpmemmed, \lpmemlarge are new to this work and are obtained using an LLM-assisted heuristic computer search \cite{caltech2026codeatlas}.

\medskip
\noindent\textbf{\lpproc\ processor code.}
The LP processor code for the balanced architecture is built from a $3 \times 5$ seed matrix $A$ over the ring \mbox{$R = \mathbb{F}_2[x]/(x^{33}+1)$}, so that $\ell = 33$. The seed matrix is
\begin{equation}\label{eq:seed_lp20}
  A = \begin{pmatrix}
    1      & 1      & 1      & 1      & 1      \\
    1      & x^{14}  & x^{19} & x^{11} & x^{26} \\
    1      & x^{13} & x^{2} & x^{15} & x^{21}
  \end{pmatrix},
\end{equation}
where we have written $x^0 = 1$. The LP construction in Eq.~\eqref{eq:lp_parameters} and numerical distance estimation yields a code with parameters
\begin{equation}
  [[n = 1122, \quad k \geq 132, \quad d \leq 20\,]].
\end{equation}
We find that the code has \mbox{$k = 148$}, exceeding the lower bound. The code has rate $k/n \approx 0.132$ and stabilizer weight \mbox{$3 + 5 = 8$}. The smaller block size and lower stabilizer weight make this code well-suited for use as a processor, where the overhead of performing logical gates scales with the code size (Appendix~\ref{app:surgery}). As shown in Extended Data Fig.~\ref{fig:processor_numerics}a, $\lpproc$ achieves an extrapolated block error rate (per code cycle) of $\lesssim 10^{-11}$ at a physical error rate of $p=0.1\%$.

\medskip
\noindent\textbf{\lpmemsmall\ memory code.}
The first memory code is built from a $3 \times 7$ seed matrix $A$ over the ring \mbox{$R = \mathbb{F}_2[x]/(x^{45}+1)$}, so that $\ell = 45$. The seed matrix is
\begin{equation}\label{eq:seed_lp16}
  A = \begin{pmatrix}
    x^{29} & x^{21} & x^{31} & x^{15} & x^{37} & x^{25} & x^{27} \\
    x^{13} & x^{25} & x^{19} & x^{26} & x^{11} & x^{18} & x^{29} \\
    x^{31} & x^{2}  & x^{27} & x^{32} & x^{41} & x^{41} & x^{18}
  \end{pmatrix}.
\end{equation}
The LP construction in Eq.~\eqref{eq:lp_parameters} and numerical distance estimation yields a code with parameters\begin{equation}
  [[n = 2610, \quad k \geq 720, \quad d \leq 16\,]].
\end{equation}
We find that the code achieves $k = 744$, slightly exceeding the lower bound guaranteed by the construction. The code has rate $k/n \approx 0.285$ and stabilizer weight $3+7 = 10$. As shown in Fig.~\ref{fig:codes}a, $\lpmemsmall$ achieves an extrapolated block error rate per cycle of $\approx$\,$10^{-7}$ at $p=0.1\%$.

\medskip
\noindent\textbf{\lpmemmed\ memory code.}
The second memory code is built from a $3 \times 7$ seed matrix $A$ over the ring \mbox{$R = \mathbb{F}_2[x]/(x^{75}+1)$}, so $\ell = 75$. The seed matrix is
\begin{equation}\label{eq:seed_lp22}
  A = \begin{pmatrix}
    x^{0}  & x^{71} & x^{73} & x^{68} & x^{33} & x^{50} & x^{47} \\
    x^{38} & x^{39} & x^{60} & x^{26} & x^{18} & x^{1}  & x^{23} \\
    x^{73} & x^{6}  & x^{5}  & x^{42} & x^{20} & x^{22} & x^{73}
  \end{pmatrix}.
\end{equation}
The LP construction in Eq.~\eqref{eq:lp_parameters} and numerical distance estimation yields a code with parameters
\begin{equation}
  [[n = 4350, \quad k \geq 1200, \quad d \leq 20\,]].
\end{equation}
We find that the code achieves $k = 1224$, slightly exceeding the lower bound guaranteed by the construction. The code has rate $k/n \approx 0.281$ and stabilizer weight $3 + 7 = 10$. As shown in Fig.~\ref{fig:codes}a, $\lpmemmed$ achieves an extrapolated block error rate per cycle of $\lesssim 10^{-10}$ at $p=0.1\%$.

\medskip
\noindent\textbf{\lpmemlarge\ memory code.}
The third memory code is built from a $3 \times 7$ seed matrix $A$ over the ring \mbox{$R = \mathbb{F}_2[x]/(x^{91}+1)$}, so $\ell = 91$. The seed matrix is
\begin{equation}\label{eq:seed_lp24}
  A = \begin{pmatrix}
    x^{57} & x^{75} & x^{42} & x^{80} & x^{7}  & x^{67} & x^{27} \\
    x^{57} & x^{73} & x^{34} & x^{12} & x^{27} & x^{50} & x^{87} \\
    x^{21} & x^{53} & x^{70} & x^{18} & x^{1}  & x^{3}  & x^{18}
  \end{pmatrix}.
\end{equation}
The LP construction in Eq.~\eqref{eq:lp_parameters} and numerical distance estimation yields a code with parameters
\begin{equation}
  [[n = 5278, \quad k \geq 1456, \quad d \leq 24\,]].
\end{equation}
We find that the code achieves $k = 1480$ logical qubits, again exceeding the lower bound. The code has rate $k/n \approx 0.280$ and stabilizer weight $3+7 = 10$. As shown in Fig.~\ref{fig:codes}a, $\lpmemlarge$ achieves an extrapolated block error rate per cycle of $\lesssim 10^{-11}$ at $p=0.1\%$.

\subsection{Bivariate bicycle codes\label{app:bb_codes}}
The BB code family~\cite{bravyi2024high} can be viewed as a special subfamily of LP codes. A BB code $\mathrm{LP}(a, b)$ is defined by two elements \mbox{$a, b \in \mathbb{F}_2[x, y]/(x^l + 1, y^m + 1)$} in the bivariate polynomial ring with periods $l$ and $m$. The resulting code acts on $n = 2lm$ physical qubits with stabilizer generators determined by the polynomials $a$ and $b$. The BB code used in this work functions as both a processor and factory code.

\begin{figure*}[t!]
    \centering
    \includegraphics{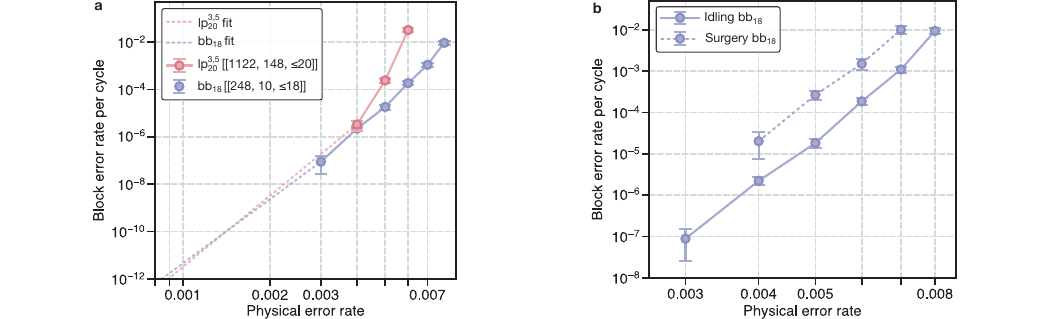}
    \caption{\textbf{Processor code and surgery performance.} \textbf{a,}~Block error rate per cycle as a function of physical error rate for the $\lpproc$ (pink) and $\bb$ (blue) codes. A least-square fit to both the $\bb$ and $\lpproc$ data yields a slope greater than $d/2$, the theoretical maximum value in the limit that the physical error rate $p\to 0$. We therefore fit the data point at the smallest physical error rate of \lpproc\ (pink dashed line) the form $y=ax^{10}$, where we find \mbox{$a= 18.5$}. Using the same procedure for \bb\ (blue dashed line) to fit a line of the form $y=ax^{9}$,  we find \mbox{$a= 15.65$}. \textbf{b,}~Block error rates per cycle versus physical error rate for the $\bb$ code, either idling for $9$ cycles (solid line) or performing a surgery measurement of a high-weight logical operator (Table~\ref{tab:surgery}) with $\taus = 15$ cycles (dashed line). Surgery failure rates are averaged over $X$- and $Z$-basis initialization and measurement experiments.
    }
    \label{fig:processor_numerics}
\end{figure*}

\medskip
\noindent\textbf{$\bb$\ processor and factory code.}
We use a BB code from Ref.~\cite{symons2025sequences} with parameters $[[248, 10, \leq 18]]$, defined by the polynomials
\begin{equation}\label{eq:bb_processor}
  a = 1 + x^6y + x^{27}, \quad b = y^2 + x^{15}y^3 + x^{24},
\end{equation}
with $l = 31$ and $m = 4$. We refer to this code as $\bb$. The code has rate $k/n = 10/248 \approx 0.04$ and stabilizer weight $6$. As shown in Extended Data Fig.~\ref{fig:processor_numerics}a, $\bb$ achieves an extrapolated block error rate (per code cycle) $\lesssim 10^{-11}$ at $p=0.1\%$. Compared to the LP processor code $\lpproc$, this code has a significantly smaller block size and comparable distance, but encodes far fewer logical qubits per block. 
The choice between these two processor codes therefore introduces a tradeoff between per-block resource cost and the number of logical qubits per block.

\begin{table}[t!]
    \caption{\textbf{Code parameters.} Details of the codes used in this work, including the code parameters $[[n, k, d]]$, the stabilizer weight, and the encoding rate $k/n$. \label{tab:all_codes}}
    \centering
    \begin{ruledtabular}
    \begin{tabular}{p{.19in}ccccc}
    Code & $n$ & $k$ & $d$ & Weight & Rate \\
    \hline
    $\bb$        & 248  & 10  & $\leq 18$ & 6 & 0.04 \\
    $\lpproc$ & 1122 & 148  & $\leq 20$ & 8 & 0.13 \\
    $\lpmemsmall$ & 2610 & 744  & $\leq 16$ & 10 & 0.29 \\
    $\lpmemmed$ & 4350 & 1224  & $\leq 20$ & 10 & 0.29 \\
    $\lpmemlarge$ & 5278 & 1480 & $\leq 24$ & 10 & 0.28 \\
    \end{tabular}
    \end{ruledtabular}
\end{table}

\section{Surgery\label{app:surgery}}
Universal computation on general qLDPC codes can be performed with logical Pauli product measurements (PPMs), which implement Clifford gates, and non-Clifford magic resource states. Code surgery~\cite{cohen2022low, cross2024improved, williamson2024low, ide2024fault} is a method for implementing addressable PPMs on general qLDPC codes. First introduced in Ref.~\cite{cohen2022low}, it generalizes lattice surgery for topological codes~\cite{horsman2012surface} to general qLDPC codes. It has since been actively developed as a practical and flexible framework for fault-tolerant logical operations~\cite{cross2024improved, williamson2024low, ide2024fault, he2025extractors, swaroop2024universal, zhang2025time, cowtan2025parallel, zheng2025high}.

Here we adapt these existing techniques to our architecture, analyze their concrete resource costs, and discuss their limitations and possibilities for further improvement. We refer readers to Sec.~3 of Ref.~\cite{he2025extractors} for a comprehensive review of general code-surgery techniques, and provide a brief summary for completeness below. 

\subsection{Description \label{app:surgery_description}}
Given a qLDPC code $\mathcal{Q}$, a surgery gadget measures a collection of logical Pauli operators $\mathcal{L}= \{\bar{P} \in \bar{\mathcal{P}}_k\}$, where $\bar{\mathcal{P}}_k$ denotes the logical Pauli group acting on the $k$ encoded qubits. This is achieved by coupling $\mathcal{Q}$ to an ancilla system $\mathcal{A}$ such that the eigenvalues of the logical operators in $\mathcal{L}$ are extracted non-destructively via measurements on $\mathcal{A}$~\cite{cohen2022low, cross2024improved, williamson2024low}. 

Here we briefly describe the procedure for measuring $X$-type logical operators. The procedures for other types of logical operators can be adapted analogously~\cite{he2025extractors, cowtan2025parallel}. For the data code $\mathcal{Q}(Q, S_X, S_Z)$, we denote by $Q$ the set of qubits, and by $S_X$ and $S_Z$ the sets of $X$- and $Z$-checks, respectively. We construct an ancilla system $\mathcal{A}(Q', S_X', S_Z')$ with qubits $Q'$ and check sets $S_X'$ and $S_Z'$. The checks $S_X$ and $S_Z$ (resp. $S_X'$ and $S_Z'$) act on $Q$ (resp. $Q'$) according to the parity-check matrices $H_X$ and $H_Z$ (resp. $H_X'$ and $H_Z'$). We then perform the following surgery procedure to measure $\mathcal{L}$:
\begin{enumerate}
    \item Initialize $Q^{\prime}$ in $\ket{0}$ states.
    \item Measure $\taus$ cycles of checks $S_X\cup S_X^{\prime} \cup S_Z \cup S_Z^{\prime}$ that are now deformed to support on the \emph{merged code} with qubits $Q\cup Q^{\prime}$: in addition to the original connection, $S_X^{\prime}$ is also connected to $Q$ and $S_Z$ is also connected to $Q^{\prime}$ according to matrices $f_X^{\prime}$ and $f_Z$, respectively, depending on the target logicals to measure.
    \item Detach $\mathcal{Q}^{\prime}$ from $\mathcal{Q}$ by measuring $Q^{\prime}$ in the $Z$ basis, followed by adaptive Pauli corrections on $\mathcal{Q}$.  
\end{enumerate}

The key step is measuring the checks of the \emph{merged code} for $\taus$ cycles (step 2). We denote the merged code $\Tilde{Q}(\Tilde{Q},\Tilde{S}_X, \Tilde{S}_Z)$ with check matrices
\begin{equation}
    \Tilde{H}_X = \left( \begin{array}{cc}
        H_X & 0 \\
        f_X^{\prime} & H_X^{\prime}
    \end{array}\right),
    \quad 
     \Tilde{H}_Z = \left( \begin{array}{cc}
        H_Z & f_Z \\
        0 & H_Z^{\prime}
    \end{array}\right).
\end{equation}
During this merging step, the information of $\mathcal{L}$ is extracted through measurements of the merged $X$-checks $S_X'$. Concretely, each $\bar{P} \in \mathcal{L}$ becomes an $X$-check in the row span of $\tilde{H}_X$. This requires $\langle \mathcal{L} \rangle = f_X^{\prime T}\ker(H_X^{\prime T})$, meaning that the ancilla system is designed such that the kernel of $H_X^{\prime T}$ maps exactly to the target logical operators of the data code. We focus on protocols in which $f_X^{\prime T}$ has the structure that a subset of qubits in $Q$ is connected one-to-one with a subset of $X$-checks in $S_X'$. In this case, the logical operators to be measured are fully specified by the kernel of the classical code $H_X^{\prime T}$. Depending on the dimension of $\ker(H_X^{\prime T})$, we classify surgery gadgets into either \emph{low-rate surgery} or \emph{high-rate surgery}, described below. The fault-tolerance of the surgery gadget is ensured if the following conditions hold~\cite{cross2024improved, he2025extractors, zheng2025high}: (i) the merged code $\tilde{\mathcal{Q}}$ has distance $\tilde d = \Theta(d)$, (ii) the merged code remains qLDPC, and (iii) $\taus = \Theta(d)$. We discuss how these criteria can be satisfied below. Note that the merged code can be a subsystem code with additional gauge qubits arising from the ancilla system $\mathcal{A}$~\cite{cohen2022low, ide2024fault, zheng2025high}, and in general we are referring to the distance of the subsystem code when we refer to the distance of the merged code $\tilde{d}$.

\emph{Low-rate surgery}: $\dim \ker H_X^{\prime T} = 1$. In this case, we always measure one logical operator $\bar{P}$ per gadget. The representative constructions~\cite{cross2024improved, williamson2024low, he2025extractors} use a connected graph $G(V, E)$, with a set of vertices $V$ representing $S_X^{\prime}$ and edges $E$ representing $Q^{\prime}$, for the ancilla system. This naturally guarantees $\dim \ker H_X^{\prime T} = 1$ since $\ker H_X^{\prime T}$ is simply the number of connected components of $G$. $S_Z^{\prime}$ is chosen to be a cycle basis of $G$. The distance of the merged code is guaranteed to be $\tilde{d} \geq d$ if the boundary Cheeger constant of the graph $G$ is $\geq 1$~\cite{swaroop2024universal}; the merged code can also be made qLDPC using graph-theoretical techniques such as cellulation and decongestion~\cite{cross2024improved}. These conditions can be satisfied using an ancillary graph of size \mbox{$|\mathcal{A}| = \mathcal{O}(|P| \log^3 |P|)$}, where $|P|$ denotes the physical Hamming weight of $\bar{P}$. Such a graph-based gadget can also be used to measure $\bar{P}$ of any type, i.e. a product of single-qubit logical Pauli $\bar{X}$, $\overline{Y}$, and $\bar{Z}$ operators. A typical approach~\cite{cross2024improved, williamson2024low}, which we adopt in this work, constructs the ancilla system $\mathcal{A}$ \emph{dynamically} depending on the target logical operator $\bar{P}$ to be measured. More modular alternatives that employ a fixed ancilla system for measuring different logical operators have also been developed. For example, the extractor construction~\cite{he2025extractors} enables such a modular design at the cost of a larger ancilla overhead of $O(n \log^3 n)$. Another approach leverages codes with rich automorphism groups~\cite{cross2024improved, webster2025explicit}, which reduces the task to measuring lower-weight ``seed'' logical operators, potentially leading to lower overhead.

\emph{High-rate surgery}: $\dim \ker H_X^{\prime T} = t > 1$. This gadget measures up to $t$ logical operators in parallel. Note that this entails that the ancilla system $\mathcal{A}$ is either a graph with multiple connected components (i.e. a union of disjoint, graph-based ancilla systems) or a hypergraph. Ref.~\cite{cowtan2025parallel}, improving upon Ref.~\cite{zhang2025time}, introduces a general scheme that can measure an arbitrary set of logical operators (including $\overline{Y}$-operators and a mixture of $\bar{X}$-type and $\bar{Z}$-type operators) of an $[[n,k,d]]$ qLDPC code with logically disjoint support with a provable bound on the ancilla size $\tilde{\mathcal{O}}(k\omega)$, where $\omega$ is the maximum physical weight of any logical Pauli being measured; Refs.~\cite{zheng2025high, cowtan2025fast, chang2026constant, gu2026qgpuparallellogicquantum} present techniques for measuring certain restricted patterns of $\bar{Z}$-type and $\bar{X}$-type logicals with an ancilla size $\mathcal{O}(n)$, i.e. bounded by the code block size, for $t = O(k)$. More generally, Ref.~\cite{zheng2025high} also presents a randomized algorithm that can measure a more flexible set of $\bar{Z}$-type or $\bar{X}$-type logicals with low overhead and numerically observes that it can measure up to $t \sim k$ logicals for a variety of codes with an ancilla size comparable to the original code size. Here, with a slight abuse of definitions, we refer to any scheme that can measure multiple logicals in parallel using an ancilla whose size is bounded by that of the data code, i.e. $|\mathcal{A}| = \tilde{O}(n)$, as a \emph{high-rate surgery} gadget. Note that, although we do not know such a construction yet that can measure an arbitrary set of logically non-overlapping logicals (including $Y$- and $XZ$-type) for any code with a guaranteed low overhead, we expect that more advanced schemes based on Refs.~\cite{zheng2025high, gu2026qgpuparallellogicquantum, chang2026constant, zheng2026high} can approach this goal, at least for certain structured codes.

So far, we have focused on surgery gadgets for a single code block. We now briefly discuss how to perform logical PPMs across multiple code blocks. In principle, one may treat several code blocks as a single combined code by taking their union and then apply the standard techniques described above. Alternatively, to measure a Pauli operator of the form $\bar{P}\bar{P}'$, where $\bar{P}$ and $\bar{P}'$ are supported on different code blocks (possibly of different code families), one can construct two independent ancilla systems $\mathcal{A}$ and $\mathcal{A}'$ for measuring $\bar{P}$ and $\bar{P}'$, respectively, and then connect them through a bridge system $\mathcal{B}$~\cite{cross2024improved, swaroop2024universal} to measure their product. Using the construction algorithm of Ref.~\cite{swaroop2024universal}, the bridge system requires as few as $d$ qubits and $d-1$ checks, where $d$ is the minimum distance of the two codes involved. When either $\mathcal{A}$ or $\mathcal{A}'$ is much larger than $d$, the additional qubit overhead of the bridge system is negligible compared to the size of the original ancilla systems.

\subsection{Concrete construction and benchmarking \label{app:surgery_concrete}}
Using the general procedure described in the previous section, we present concrete constructions of surgery gadgets for the space-efficient and balanced architecture in this section. Recall that these architectures consist of a $[[\nm, \km, \dm]]$ memory code, a $[[\nproc, \kproc, \dproc]]$ processor code, and several $[[\nf, \kf, \df]]$ factory codes (see Fig.~\ref{fig:codes}b and Extended Data Table~\ref{tab:notation}). As discussed in Appendix~\ref{app:compilation}, the logical PPMs required for computation have the following forms: $\bar{Z}_i \bar{P}$ or $\bar{P}$, where $\bar{Z}_i$ is a single-qubit logical Pauli $\bar{Z}$ operator acting on either the memory or a factory code, and $\bar{P}$ is an arbitrary $\kproc$-qubit logical Pauli operator on the processor code; or $\bar{X}_j$, a single-qubit logical Pauli $\bar{X}$ operator acting on either the memory or a factory code.

Accordingly, we introduce three ancilla systems $\Am$, $\Aproc$, and $\Af$ for measuring logical Pauli operators on the memory, processor, and factory codes, respectively. In addition, two bridge systems are used to connect $\Am$ and $\Aproc$, and $\Aproc$ and $\Af$, respectively. These PPMs are implemented using \emph{low-rate surgery} techniques (see the previous section) such that one logical operator of the above form is measured at a time using a single surgery gadget.

Using standard surgery constructions~\cite{cross2024improved, williamson2024low, zheng2025high}, which dynamically generate ancilla systems depending on the target logical operator, the number of distinct ancilla systems scales as $O(k_m)$, $O(k_f)$, and $O(\exp(k_p))$ in the worst case for $\Am$, $\Af$, and $\Aproc$, respectively. In architectures with fixed connectivity, such as superconducting circuits~\cite{yoder2025tour}, each distinct ancilla system would need to be physically constructed and connected to the data codes, significantly increasing the required hardware footprint. In contrast, by leveraging the reconfigurable connectivity of atom arrays, it suffices to reserve an operation zone with enough qubits to accommodate the largest ancilla configuration, namely $\max_{\Am,\Aproc,\Af} \left(|\Am| + |\Aproc| + |\Af|\right)$ (note that here we do not include ancilla qubits for the bridge systems, which are subleading in size). The ancilla systems can then be dynamically reconfigured within this region as needed. We note that this space saving comes at the cost of increased overhead in classical processing and control, as one must construct the distinct surgery gadgets in real-time and adaptively program the corresponding ancilla configurations. We discuss possible approaches to reducing these classical overheads and further improving resource efficiency in Appendix~\ref{app:surgery_discussion}.

As summarized in Extended Data Table~\ref{tab:surgery}, we construct and benchmark representative instances of the ancilla systems described above. The first and last rows present $\Am$ and $\Af$, used to measure a single-qubit logical $X$ operator on the $\lpmemmed$ and $\lpmemlarge$  memory codes and the $\bb$ factory code, respectively. Minimizing the size of these ancilla systems requires selecting a logical basis composed of conjugate pairs of single-qubit logical operators with low physical weight. For the $\bb$ code, we numerically identify such a basis in which nine logical qubits have weight-$18$ logical operators, while one has a weight-$20$ logical operator. Finding a comparable basis for the larger memory LP codes is more challenging numerically. Upcoming work~\cite{zheng2026high} provides an algebraic construction of a low-weight basis for LP codes, in which the $1{,}220$ (resp. $1{,}476$) logical qubits in $\lpmemmed$ (resp. $\lpmemlarge$) have physical weight at most $200$ (resp. $224$). Furthermore, by exploiting code automorphisms, measuring any of the above low-weight single-qubit logical $\bar{X}$ operators of $\lpmemmed$ (resp. $\lpmemlarge$) requires only $112$ (resp. $144$) distinct surgery gadgets~\cite{zheng2026high}. We report, in the first and last rows of Extended Data Table~\ref{tab:surgery}, the gadgets that measure the maximum-physical-weight single-qubit logical $\bar{X}$ operators for $\lpmemmed$, $\lpmemlarge$ and $\bb$, respectively. The middle rows present instances of $\Aproc$ used to measure high-weight logical $\bar{X}$ operators on the $\bb$ and $\lpproc$ processor codes. Each operator is selected as the maximum-physical-weight example among $10^5$ randomly sampled logical multi-qubit $\bar{X}$ operators, with logical (resp. physical) weights $11$ (resp. $104$) for $\bb$, and $69$ (resp. $460$) for $\lpproc$.

All ancilla systems are constructed using standard graph-based surgery techniques~\cite{cross2024improved, williamson2024low}, which will be further detailed in Ref.~\cite{zheng2026high}. The distances of the surgery gadgets are estimated using the QDistRnd package~\cite{pryadko2023qdistrnd} with at least $100{,}000$ trials. Extended Data Figure~\ref{fig:processor_numerics}b shows the logical error rates of the surgery gadget listed in Extended Data Table~\ref{tab:surgery} that measures a high-weight logical operator on the $\bb$ code, corresponding to $\Aproc$ in the space-efficient architecture. With $\taus = 15 \approx 2d/3$, the surgery error rates are within an order of magnitude of the $\bb$ block failure rate at idling, demonstrating fault tolerance of the surgery construction. We expect similar behavior for the larger gadgets in Extended Data Table~\ref{tab:surgery}, and leave detailed simulations to future work.

Although Extended Data Table~\ref{tab:surgery} benchmarks ancilla systems for measuring $\bar{X}$-type logical operators, the same constructions readily extend to other logical types (e.g., $\overline{Y}$) by modifying how the ancilla systems couple to the data codes~\cite{he2025extractors}. 
Accordingly, we estimate the space cost of the ancilla systems in our architecture (i.e., the size of the operation zone), $\max_{\Am,\Aproc,\Af} \bigl(|\Am| + |\Aproc| + |\Af|\bigr)$, based on the examples in Extended Data Table~\ref{tab:surgery}.

\begin{table*}
    \caption{\textbf{Surgery systems.} Ancilla systems for measuring the relevant PPMs on different zones and codes.
    For each ancilla system, we report the target logical operators it measures and their logical weights, the number of ancillary qubits, $X$ stabilizers, and $Z$ stabilizers, as well as the resulting degree and the distance upper bound of the merged code.
    \label{tab:surgery}
    }
    \centering
    \begin{tabular}{c|c|c|c|c|c}
    \hline
    \hline
    Zone & Codes & PPMs & (Qubits, $X$-checks, $Z$-checks) & Degree & Distance  \\
    \hline
    \multirow{2}{*}{Memory} & 
    $\lpmemmedparams$-$\lpmemmed$ & 
    $\bar{P}, |\bar{P}|=1$ & ($342$, $200$, $143$) & $12$ & $\leq 20$ \\
    & $\lpmemlargeparams$-$\lpmemlarge$ & 
    $\bar{P}, |\bar{P}|=1$ & ($364$, $208$, $157$) & $12$ & $\leq 22$ \\
    \hline
    \multirow{2}{*}{Processor} 
    & $\bbparams$-$\mathsf{bb}_{18}$ & 
    $\bar{P}, |\bar{P}|=9$ & ($189$, $104$, $86$) & $9$ & $\leq 18$ \\
    & $\lpprocparams$-$\lpproc$ & 
    $\bar{P}, |\bar{P}|=69$ & ($813$, $460$, $357$) & $10$ & $\leq 20$ \\
    \hline
    Resource & $\bbparams$-$\mathsf{bb}_{18}$ & 
    $\bar{P}, |\bar{P}|=1$ &  ($39$, $20$, $20$) & $7$ & $\leq 18$ \\
    \hline
    \hline
    \end{tabular}
\end{table*}

\subsection{Resource costs and future improvements \label{app:surgery_discussion}}
We now summarize the time and space costs of the surgery instructions used in this architecture.
Based on the examples in Extended Data Table~\ref{tab:surgery} as well as the analysis in the previous sections, we estimate that the ancillary systems required to perform all relevant surgery operations in our architecture occupy $\Na = 894$ and $\Na = 1{,}874$ (resp. $\Na = 924$ and $\Na = 1{,}904$) qubits for the space-efficient and balanced architectures, respectively, when using the $\lpmemmed$ (resp. $\lpmemlarge$) memory code (see also Extended Data Table~\ref{tab:space_breakdown}). We assume that each surgery gadget is implemented in $\taus \approx 2d/3$ code cycles, where $d$ denotes the distance of the processor code. This choice balances the space-like and time-like logical error rates of the gadget, thereby approximately minimizing the overall logical error rate. See Sec.~\ref{app:surgery_concrete} for numerical justification of this choice.

There are several avenues to reduce the classical overhead associated with constructing and implementing surgery gadgets, as well as further lowering the overall resource costs. First, the number of distinct surgery gadgets could be reduced using extractor-style constructions~\cite{he2025extractors}, or more modular approaches that rely on a small set of fixed gadgets and bridge systems by exploiting symmetries of the data code, particularly automorphisms~\cite{cross2024improved, webster2025explicit, webster2026pinnacle}. For codes with sufficiently rich symmetries, the latter approach may enable a fixed ancilla system that is even smaller than the data block while still supporting measurements of arbitrary logical operators~\cite{webster2025explicit}. In addition, alternative techniques beyond code surgery may further reduce overhead, including homomorphic measurements~\cite{huang2022homomorphic, xu2025fast, yang2026rascql} and in-block transversal gates~\cite{malcolm2025computing, yang2026rascql, sayginel2025fault}. Some of these directions will be explored in upcoming work~\cite{zheng2026high} for the lifted-product code family. 

\section{Magic\label{app:magic}}

Universal quantum computation requires non-Clifford gates in addition to Clifford gates; typically, $\overline{T}$ or $\overline{\text{CCZ}}$ gates are used. Non-Clifford gates can be executed by generating magic resource states (e.g., \mbox{$\lket{T}=\overline{T}\lket{+}$} and \mbox{$\ccz = \overline{\text{CCZ}}\lket{+}^{\otimes 3}$}) and performing gate teleportation with conditional Clifford feedforward gates~\cite{bravyi2005universal}. In this work, we implement Toffoli gates by distilling high-fidelity $\ccz$ states and teleporting them into the computation using logical PPMs~\cite{litinski2019magic, litinski2022active}.

Historically, generating magic resource states has been among the most resource-demanding subroutines for implementing large-scale logical algorithms for planar surface-code architectures~\cite{gidney2019how, litinski2019magic}. Recent techniques such as magic state cultivation~\cite{gidney2024magic} substantially reduce the cost of generating magic states in planar architectures. However, cultivation uses surface codes with low encoding rates, and cultivation alone cannot always achieve the logical error rates required for large-scale algorithms~\cite{gidney2025how}. Here, we describe a protocol that generates many magic states in parallel with low overhead and logical error rates by combining surface-code cultivation with high-rate codes.

\subsection{Description \label{app:magic_description}}
\begin{figure}
\includegraphics[width=1\linewidth]{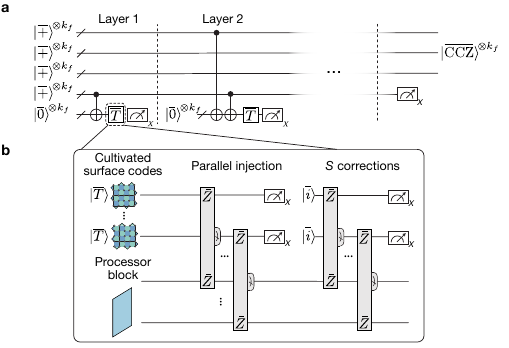}
    \caption{\textbf{Parallel magic state distillation.} We distill $\kf$ $\ccz$ states using five copies of $[[\nf, \kf, \df]]$ factory codes. Low-error cultivated surface-code $\lket{T}$ states are loaded into one processor code using parallel surgery~\cite{zhang2025constant}. Transversal CNOT gates between this code and the four remaining processor codes are used to perform the PPMs in 8$T$-to-CCZ distillation~\cite{litinski2019magic}, outputting three blocks of high-fidelity $\ccz$ states. 
    }
    \label{fig:distillation}
\end{figure}

As illustrated in Extended Data Fig.~\ref{fig:distillation}, the high-rate magic state distillation procedure distills cultivated $\lket{T}$ states on small surface codes into high-fidelity $\ccz$ states hosted in three high-rate factory code blocks with parameters $[[\nf, \kf, \df]]$, using the 8$T$-to-CCZ distillation circuit in Ref.~\cite{litinski2019magic}. The distillation circuit has eight layers of gates, each with a similar structure. For a given layer, we use an ancillary factory code block, denoted as the $\overline{T}$ block, to implement parallel logical Pauli product rotations on the remaining four processor code blocks (see Fig.~16 of Ref.~\cite{litinski2019magic} for the concrete circuit). This is done using up to four transversal CNOT gates between the $\overline{T}$ block and the remaining processor code blocks, followed by a transversal faulty $\overline{T}$ measurement~\cite{litinski2019magic}.

As shown in Extended Data Figure~\ref{fig:distillation}b, the transversal faulty $\overline{T}$ measurements on the $\overline{T}$ block are implemented by parallel teleportation of $\kf$ cultivated $\lket{T}$ states, each hosted in a small $[[\ns, \ks, \ds]] = [[\ds^2, 1, \ds]]$ surface-code patch. For the parallel teleportation we use parallel code surgeries~\cite{zhang2025constant} (Appendix~\ref{app:surgery}), followed by local $\overline{S}$ corrections via teleportation of $\lket{Y}$ states. The $\lket{Y}$ states can be prepared using a subset of the patches used for cultivation, for example using fold-transversal gates in depth one~\footnote{Note that one could also teleport auto-corrected surface code $\lket{T}$ states, which avoids the $\overline{S}$ corrections, at the price of using more surface code patches~\cite{litinski2019magic}.}. Finally, we perform transversal $X$-basis measurements on the $\overline{T}$ block, followed by Pauli feedback on the other four factory blocks. This procedure is repeated eight times, once for each layer of logical Pauli-product rotations in Fig. 16 of Ref.~\cite{litinski2019magic}. The resulting circuit generates $\kf$ $\ccz$ states uniformly distributed among three factory code blocks, i.e. $\ccz^{\otimes \kf}$. The correctness of the generated magic states is heralded by the fourth factory block being measured successfully in $\lket{+}^{\otimes \kf}$ in a transversal $X$-basis measurement.

Using factory codes with a sufficiently large distance \mbox{$\df \gg \ds$}, noise on the Clifford operations such as transversal CNOT gates can be neglected to good approximation~\cite{litinski2019magic}, and the main faulty operations in the circuit are the $\overline{T}$ gates on the $\overline{T}$ block via teleporting the surface code $\lket{T}$ states. As such, the logical error rates of the output $\ccz$ states depend primarily on the error rates of these noisy $\overline{T}$ gates, which depend on the fidelity of the cultivated $\lket{T}$ states. In addition, there are also extra errors introduced to the $\overline{T}$ factory code block when performing the parallel surgeries between the high-distance factory code and the low-distance surface codes. For example, in order to avoid introducing logically correlated errors on the factory code due to low weight physical faults during the surgery, the protocol in Ref.~\cite{zhang2025constant} requires a surgery ancilla system with size $\tilde{O}(\kf \df)$ and $O(\df)$ code cycles, which exponentially suppresses the correlated error by $\df$, where $\tilde{O}(\cdot)$ indicates scaling up to logarithmic factors. However, because we are performing transversal distillation across different high-rate blocks rather than within a high-rate code block, we can allow for correlated errors within the $\overline{T}$ block. This is because distillation across blocks effectively involves $\kf$ parallel distillation factories.

As such, we are only concerned about the \emph{marginal error rate} of each $\overline{T}$ gate in a $\overline{T}$ factory code block. For this, it is sufficient to use surgery ancillas of size $\tilde{O}(\kf \ds)$ and $\tauscult = O(\ds)$ code cycles, which exponentially suppresses the extra (possibly correlated) errors during the surgery process by $\ds$. We assume that each layer of surgery takes $\tauscult = 2\ds/3$ code cycles, which often minimizes the total logical error rates~\cite{cross2024improved}, and the final marginal error rate for the $\overline{T}$ gate is approximately $2 p_T$, where $p_T$ is the error rate of the cultivated surface code $\lket{T}$ states. We then estimate that the logical error rate of each output $\ccz$ state is $p^L_{\mathrm{CCZ}}  \approx 28(2 p_T)^2$~\cite{litinski2019magic}. The success probability of the protocol scales linearly with $p_T$. Since we typically have $p_T \ll 1$, the success probability that the entire block succeeds is very close to $1$.

The total time cost is approximately \mbox{$8\times (\tau_T + 2\tauscult)$}, where $\tau_T$ denotes the time to cultivate each surface code $\lket{T}$ state. As for space cost, the surgery ancilla is typically much smaller than the ancilla required for performing fully fault-tolerant logic for other modules of our architecture, e.g. the processor code or the memory code, so we neglect it here and do not include it into the space cost of the operation zone. This can also be justified if we can pipeline the surgery operations across the full architecture such that a subset of the ancillary qubits in the operation zone can be reconfigured for executing the surgeries in the factory. As such, and based on the analysis above, we estimate that the space cost of the entire factory is \mbox{$5 \Nf + \kf \Ns$}.

\subsection{Resource costs \label{app:magic_resource}}
Here we compute the concrete resource costs for the magic factory described in the main text for the space efficient and balanced architectures. For the factory code, we select the $\bb$ code (Extended Data Table~\ref{tab:all_codes}) with $\Nf= 367$. We choose the distance of the surface code to be $\ds = 7$ so that it has logical error rate $\lesssim 10^{-6}$~\cite{ismail2025transversal}. We consider the scheme in Ref.~\cite{sahay2025fold} for cultivating $\lket{T}$ states with $p_T \approx 10^{-6}$ at physical error rates of $0.1\%$. This approach involves injecting a noisy $\lket{T}$ state with fault distance 1 into a small \mbox{$d=3$} rotated surface code, then measuring the fold-transversal Hadamard operator twice (each measurement followed by a stabilizer measurement cycle). Finally, the code is grown from \mbox{$d=3$} to \mbox{$d=5$} using a unitary growth procedure, then to \mbox{$d=7$} using stabilizer measurements. Postselection is applied between the measurement and growth stages, then after the growth stage. For physically motivated noise models (see Fig.~3 of Ref.~\cite{sahay2025fold}), in expectation fewer than two total attempts are required for the cultivated state to pass all postselection tests to reach error rates of $\lesssim10^{-6}$. Note that our choice of final code distance \mbox{$\ds= 7$} is likely conservative because $\gtrsim50\%$ of errors are atomic loss in realistic settings, and such heralded errors both improve the threshold by a factor of \mbox{$\approx 2\times$} and enable better postselection~\cite{vaknin2025efficient}.

Because the injection and cultivation stages occur at a smaller code size, extra surface code blocks can be prepared in parallel such that the success probability of this stage is close to one, without exceeding the maximum space cost after growth. The total depth, from injection to growth, is approximately four cycles: one cycle for injection, one for cultivation, and two for growth. The full process requires fewer than two attempts, $\approx1.25$, in expectation. As such, we estimate that cultivating each $\lket{T}$ state takes $\tau_T \approx 5$ code cycles, and each surgery layer takes $\tauscult \approx 2\ds/3 \approx 5$ stabilizer measurement cycles. With these choices, our factory produces \mbox{$\kf = 10$} $\ccz$ states, each with an error rate $p^L_{\mathrm{CCZ}} \approx 10^{-10}$, using $2{,}565$ total qubits in $8\times(5+2\cdot 5)\approx 120 = 6\dproc$ cycles, for $\dproc = 20$ (see also Extended Data Table~\ref{tab:space_breakdown}). Note that the time cost per $\ccz$ state is $120/\kf = 12 < \dproc$. As such, in our space-efficient or balanced architecture, where we consume these $\ccz$ states sequentially, we will neglect the time cost for producing each $\ccz$ state. 

\section{Numerical simulations\label{app:numerics}}
We numerically assess the performance of our fault-tolerant architecture under circuit-level noise using Stim~\cite{gidney2021stim}. We focus on simulating the performance of the codes used in our architecture under repeated stabilizer measurements, as well as the surgery gadgets presented in Appendix~\ref{app:surgery}.

\paragraph{Noise model.}
We use a circuit-level depolarizing noise model parameterized by a single physical error rate $p$, consistent with prior work~\cite{xu2024constant, menon2025magic, li2025transversal}. The noise channels are applied as follows.
\begin{itemize}
    \item \emph{Two-qubit gate noise.} Each CNOT gate is followed by a two-qubit depolarizing channel with error rate $p$, which applies each of the 15 nontrivial two-qubit Pauli operators with equal probability $p/15$.
    \item \emph{State preparation noise.} Each state preparation ($\ket{0}$ or $\ket{+}$) is followed by a single-qubit depolarizing channel with error rate $p$. 
    \item \emph{Measurement noise.} Each measurement is preceded by a single-qubit depolarizing channel with error rate $p$.
\end{itemize}

\paragraph{Syndrome extraction circuit.}
For each code, we construct a syndrome extraction circuit that measures the $X$-type and $Z$-type stabilizers separately. For each stabilizer type, we determine the gate scheduling, specifically the ordering of CNOT gates at each time step, using the coloration scheduling method~\cite{tremblay2022constant}. This approach performs an edge coloring of the bipartite Tanner graph associated with the check matrix, ensuring that no data qubit participates in more than one gate per time step. As a result, the circuit requires $\Delta_X + \Delta_Z$ layers of CNOT gates for a code whose $X$ and $Z$-check matrices have degrees $\Delta_X$ and $\Delta_Z$, respectively. We use this generic scheduling throughout both the memory experiments and the surgery experiments described below whenever stabilizers of a (possibly deformed) quantum code need to be measured. We adopt this generic coloration scheduling primarily for ease of simulation. In practice, more structured approaches could be used, such as the ``product-coloration'' scheduling for LP codes~\cite{xu2024constant}, translationally invariant scheduling for BB codes~\cite{bravyi2024high, viszlai2023matching} that is directly compatible with current neutral-atom devices, or shorter-depth schedules that interleave the measurements of $X$ and $Z$-checks~\cite{xu2024constant, kang2025quits}. 

Here we give a detailed step-by-step of the implementation of the circuit~\cite{tremblay2022constant, xu2024constant, gidney2021stim}. First, the $X$-type ancillas are initialized in the $\ket{+}$ state via a Hadamard gate, followed by preparation noise. A sequence of CNOT gates is then applied between each $X$-ancilla (as control) and its supported data qubits (as targets), with the gate schedule determined by the edge coloring. Next, the $Z$-type ancillas are initialized in the $\ket{0}$ state, followed by preparation noise, and CNOT gates are applied with each data qubit as control and the corresponding $Z$-ancilla as target. Finally, all ancillas are measured and reset.

\paragraph{Memory experiment.}
We simulate a quantum memory experiment consisting of $d/2$ stabilizer measurement cycles, where $d$ is the code distance. The logical qubits are initialized in the $\ket{+}^{\otimes k}$ state via transversal $X$-basis preparation of all data qubits. In the first cycle, detectors are defined by the raw $X$-syndrome outcomes. In subsequent cycles, detectors are defined by the difference between consecutive $X$-syndrome outcomes, which flags any change due to errors occurring between cycles. After the final cycle, all data qubits are measured in the $X$ basis with measurement noise applied. The final-cycle detectors compare the data qubit measurement outcomes against the last cycle by reconstructing the $X$-stabilizer values from the data. Logical observables are defined by the logical $\bar{X}$ operators of the code.

\paragraph{Surgery experiment. }
We simulate a surgery gadget that measures a set of $t$ logical $\bar{X}$ operators $\mathcal{L} = \{\bar{P}_i\}_{i=1}^t$, following the general three-step protocol described in Sec.~\ref{app:surgery}. Recall that a surgery gadget consists of a $k$-logical-qubit data code with physical qubits $Q$ together with an ancilla system with physical qubits $Q'$. During the protocol, the stabilizer checks of the merged code are measured for $\taus$ cycles. For each gadget we perform two experiments. In both experiments the ancilla qubits $Q'$ are initialized and measured in the $Z$ basis, while the data qubits $Q$ are initialized and measured in either the $X$ or the $Z$ basis.

In the $X$-basis experiment, we define $k+t$ logical observables: the final logical $\bar{X}$ operators of the data code, obtained from the parities of the final single-qubit $X$ measurements on $Q$, together with the outcomes of the target logical operators $\mathcal{L}$, extracted from the parities of the merged-code $X$-checks in the first stabilizer measurement cycle. Failures of these observables characterize space-like logical $\bar{Z}$ errors (determined by the $Z$ distance of the merged code) as well as time-like logical errors corresponding to incorrect measurement of the target logical operators. In the $Z$-basis experiment, we define $k-t$ logical observables corresponding to the logical $\bar{Z}$ operators of the data code that commute with $\mathcal{L}$. Failures of these observables characterize space-like logical $\bar{X}$ errors, determined by the $X$ distance of the merged code.

\paragraph{Decoding.}
Decoding is performed using an ensemble of belief propagation with localized statistics decoders (BP-LSD)~\cite{hillmann2024localized}, implemented via the \texttt{ldpc} package~\cite{Roffe_LDPC_Python_tools_2022}. Each decoder instance uses min-sum belief propagation with serial scheduling for $100$ iterations, using the default annealing schedule for the min-sum scaling factor provided by the \texttt{ldpc} package $(\texttt{ms\_scaling\_factor} = 0.0)$. The LSD is performed using an exhaustive search of order~$5$ $(\texttt{lsd\_method} = \texttt{lsd\_e}, \texttt{lsd\_order} = 5)$. To improve decoding performance, we run an ensemble of five decoder instances per syndrome, each operating under a different channel probability model: (i)~the nominal error probabilities $\vec{p}$, (ii)~a randomized serial schedule with the nominal probabilities, (iii)~optimistic probabilities $0.8\,\vec{p}$, (iv)~pessimistic probabilities $1.2\,\vec{p}$, and (v)~thermally perturbed probabilities $\vec{p} \circ (1 + \vec{\eta})$, where $\eta_i \sim \mathcal{N}(0, 0.04)$ independently for each error mechanism. For each syndrome, all five decoders produce a candidate error vector. Candidates that satisfy the syndrome are retained, and the one with the lowest log-likelihood ratio weighted cost
\begin{equation}\label{eq:llr_cost}
  C(\vec{e}) = \sum_i \left|\log \frac{1 - p_i}{p_i}\right| \cdot e_i
\end{equation}
is selected, where $p_i$ is the prior error probability of the $i$-th error mechanism and $e_i \in \{0,1\}$ indicates whether that mechanism is included in the correction. If no candidate satisfies the syndrome, a logical failure is recorded. The decoder is designed by an LLM-assisted heuristic computer search~\cite{caltech2026codeatlas}.

\paragraph{Logical error rate.}
We report the block logical error rate: a failure is recorded whenever the selected correction, composed with the true error, acts nontrivially on any logical observable.

\section{Space-efficient and balanced architectures \label{app:compilation}} 
In this section, we present a general compilation scheme for implementing any Clifford + Toffoli circuit on our space-efficient and balanced architectures based on the logical instructions in Appendices~\ref{app:surgery} and~\ref{app:magic}. We then apply this general compilation strategy to two algorithmic subroutines---ripple-carry adders~\cite{vedral1996quantum, cuccaro2004new, gidney2018halving} and unary lookup tables~\cite{babbush2018encoding}---which are the key elementary building blocks of the cryptographic applications considered in this work, and estimate their time costs. Finally, we estimate the runtimes for \rsa\ and \ecc\ using our space-efficient and balanced architectures.

Our space-efficient or balanced architectures consist of:
\begin{itemize}
    \item A $[[\nm, \km, \dm]]$ memory code, where $\km$ is larger than the logical footprint of the entire algorithm.
    \item A $[[\nproc, \kproc, \dproc]]$ processor code.
    \item Three $[[\nf, \kf, \df]]$ factory codes hosting $\kf$ copies of logical $\ccz$ states distilled using ancillary factory code blocks, where each $\ccz$ state is encoded across codes.
    \item Additional ancillary codes for high-rate distillation, including two $[[\nf, \kf, \df]]$ factory codes and $\kf$ $[[\ns, 1, \ds]]$ surface codes for cultivation.
\end{itemize}

\begin{figure}[t!]
    \centering
    \includegraphics[width=1\linewidth]{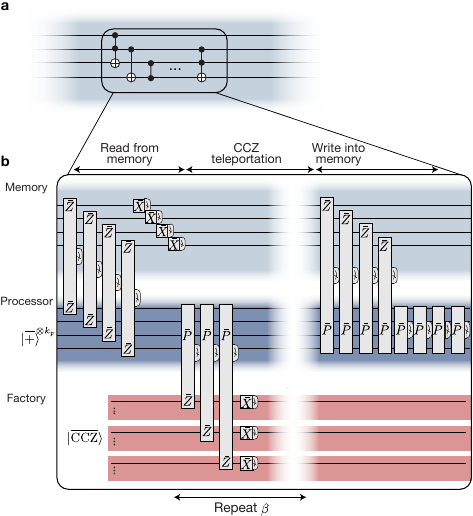}
    \caption{\textbf{Compilation strategy.} \textbf{a,} A general Clifford+Toffoli circuit is divided into $m_i$-qubit subcircuits $C_i$ containing $\beta_i$ Toffoli gates, $\gamma_i$ mid-circuit Pauli measurements, and arbitrary Clifford gates. \textbf{b,} Each sub-circuit is implemented by first teleporting the $m_i$ qubits from the memory code to the processor code. Each Toffoli gate via $\ccz$ teleportation and each mid-circuit Pauli measurement is implemented through sequential Clifford-transformed PPMs. Finally, the $m$ qubits are teleported back to the memory code, again via Clifford-transformed PPMs. The PPMs can involve high-weight logical Pauli operators on the processor code, which we denote as $\overline{P}$ for simplicity. 
    }
    \label{fig:low_rate_compilation}
\end{figure}

Given a general Clifford + Toffoli circuit on $\km$ qubits, we serialize it into sub-circuits $\{C_i\}$, where each $C_i$ is a $m_i$-qubit circuit containing $\beta_i$ Toffoli gates, $\gamma_i$ mid-circuit Pauli measurements, and arbitrary Clifford gates. We require that each sub-circuit can fit inside the processor code block, i.e. $m_i \leq \kproc$ for all $C_i$. We use $\bar{P}$, $\bar{P}^{\prime}$, and $\bar{P}^{\prime\prime}$ to indicate a logical Pauli operator of the memory code, the processor code, and the factory code, respectively. Then, as illustrated in Extended Data Figure~\ref{fig:low_rate_compilation}, we implement each $C_i$ with the following steps:
\begin{enumerate}
    \item Teleport the $m_i$ logical qubits in the memory code to the processor code by sequentially measuring $\{\bar{Z}_{I_j} \bar{Z}^{\prime}_{j}\}_{j \in [m_i]}$ and then $\{\bar{X}_{I_j}\}_{j \in [m_i]}$. Here, $I_j$ denotes the index of the $j$-th qubit among the $m_i$ qubits in the memory code that $C_i$ is supported on.
    \item Perform a $m_i$-qubit Pauli-based computation on the processor code, where Cliffords are pushed to the end and each Toffoli gate is implemented by three $\bar{P}^{\prime} \bar{Z}^{\prime \prime}$ measurements between the processor code and one of the factory codes, where $\bar{P}^{\prime}$ is a $m_i$-qubit Pauli operator. In addition, each mid-circuit Pauli measurement will be transformed to a high-weight PPM and also executed sequentially.
    \item Teleport the $m_i$ qubits back into the memory code by sequentially performing $\bar{Z}_{I_j} \bar{P}_j^{\prime}$ between the processor and the memory code, followed by $m_i$ $m_i$-qubit measurements on the processor code.
\end{enumerate}
The above protocol extensively utilizes two key subroutines: 
\begin{enumerate}
    \item \textit{Measurement-based teleportation:} logical qubit $i$ is teleported to logical qubit $j$ (potentially across different codes) by preparing $\lket{+}_j$, performing a $\bar{Z}_i \bar{Z}_j$ PPM followed by a $\bar{X}_i$ PPM, and applying the corresponding logical Pauli corrections (see Fig.~3 of Ref.~\cite{poulsen2017fault}).
    \item $\ccz$ \textit{teleportation:} given a $\ccz$ resource state on logical qubits $a$, $b$, and $c$, a $\overline{\mathrm{CCZ}}$ gate can be implemented on logical qubits $a'$, $b'$, and $c'$ by performing three PPMs $\{\bar{Z}_{a}\bar{Z}_{a'}, \bar{Z}_{b}\bar{Z}_{b'}, \bar{Z}_{c}\bar{Z}_{c'}\}$, followed by three PPMs $\{\bar{X}_a, \bar{X}_b, \bar{X}_c\}$ and the corresponding $\overline{\mathrm{CZ}}$ corrections (see Fig.~15 of Ref.~\cite{litinski2022active}).
\end{enumerate}
When incorporated into the above compilation scheme, Clifford circuits generally transform the processor-code logical operators appearing in these PPMs into higher-weight logical operators, eventually yielding the circuit form illustrated in Extended Data Fig.~\ref{fig:low_rate_compilation}.

We refer to the above execution of each $C_i$ as a \emph{computation unit}. 
Each computation unit involving a sub-circuit $C_i$ thus takes time (in units of code cycles)
\begin{equation}
    \tau(C_i)  = (4m_i + 4\beta_i + \gamma_i)\taus,
    \label{eq:time_general}
\end{equation}
where $\taus$ again denotes a surgery cycle consisting of $\taus \approx 2d_p/3$ code cycles (see Sec.~\ref{app:surgery}).

In the above, we assumed that for each $C_i$ the involved $m_i$ qubits are read from the memory at the beginning and written back to the memory at the end. In other words, the processor code is initialized and measured transversally in the $X$ basis for each computation unit. In practice, for instance, when two consecutive sub-circuits have overlapping qubit support, one could instead read only $m_i^{(\mathrm{in})} < m_i$ qubits from the memory and write back only $m_i^{(\mathrm{out})} < m_i$ qubits, leaving some logical qubits initialized or stored in the processor code across computation units. In this case, a similar compilation strategy applies, except that the Clifford frames must be tracked more carefully across different computation units. Importantly, since the I/O entangling operations between the memory and the processor are executed explicitly, the Clifford frame on the processor code remains unentangled from that of the memory, analogous to the situation in Ref.~\cite{webster2026pinnacle}. Consequently, the PPMs involve the large memory code (with single-qubit logical $\bar{Z}$ operators) only during the I/O operations. The time cost for such a computation unit with ``partial'' I/O operations is given by $\tau(C_i) = (2m_i^{(\mathrm{in})} + 2m_i^{(\mathrm{out})} + 4\beta_i + \gamma_i)\taus$.
Equivalently, one may still use Eq.~\ref{eq:time_general} by replacing $m_i$ with the amortized input--output qubit count $(m_i^{(\mathrm{in})} + m_i^{(\mathrm{out})})/2$.

In the following section, we compute the space cost for this architecture using the concrete code instances discussed in Appendix~\ref{app:codes}. Then, based on Eqs.~\eqref{eq:time_general}, we provide a concrete estimate for the runtime of ripple-carry adders and unary lookup tables, parameterized by their key parameters such as bit size. These subroutines form the core of Shor's algorithm both for ECC and RSA~\cite{jones2012layered, fowler2012surface, ogorman2017quantum, gheorghiu2019benchmarking, gidney2019how, dallairedemers2025brace, gidney2025how, zhou2025resource, webster2026pinnacle, chevignard2025reducing, proos2003shor, roetteler2017quantum, haner2020improved, litinski2023compute, gouzien2023performance, babbush2026quantum}.

\begin{table*}[t!]
\caption{\textbf{Space costs.} Breakdown of the physical qubit counts in the space-efficient and balanced architectures in different functional zones.
\label{tab:space_breakdown}
}
\centering
\begin{ruledtabular}
\begin{tabular}{ccccc}
Zone & Space-efficient, & Space-efficient, & Balanced  & Balanced\\
 & \lpmemmed memory & \lpmemlarge memory & \lpmemmed memory & \lpmemlarge memory\\

\hline
Memory & 5{,}913 & 7{,}177 & 5{,}913 & 7{,}177 \\
Processor & 367 & 367 & 1{,}609 & 1{,}609 \\
Resource & 2{,}565 & 2{,}565 & 2{,}565 & 2{,}565 \\
Operation & 894 & 924 & 1{,}874 & 1{,}904
 \\
\hline
Total & 9{,}739 & 11{,}033 & 11{,}961 & 13{,}255
 \\
\end{tabular}
\end{ruledtabular}
\end{table*}

\subsection{Space cost \label{app:space_cost}}
The space cost can be computed by summing the number of physical qubits in each of the four zones. We define the qubit footprint for a $[[n, k, d]]$ code to include data qubits as well as one basis ($X$ or $Z$) of ancilla qubits for measuring stabilizers, i.e. \mbox{$N\simeq n+\lfloor(n-k)/2\rfloor$} (Table~\ref{tab:notation}). Note that in reconfigurable systems, code stabilizers can be measured in smaller batches, reducing the qubit overhead but increasing idling times. Recall that the resource zone space cost includes five \bb\ factory blocks with parameters \mbox{$[[\nf, \kf, \df]]=\bbparams$}, and $\kf=10$ \mbox{$[[\ns, \ks, \ds]]=[[49,1,7]]$} cultivated surface codes. The operation zone costs are derived from Extended Data Table~\ref{tab:surgery}, using only one basis of ancilla qubits. The resulting qubit counts are listed in Extended Data Table~\ref{tab:space_breakdown}.

\subsection{Time cost for adders\label{app:adders}}
Here we estimate the time costs of performing addition in the space-efficient and balanced architectures. An adder has action
\begin{equation}
    \ket{a}\ket{b} \to \ket{a}\ket{a+b}
\end{equation}
on two $n$-bit numbers \mbox{$\ket{a} = \ket{a_1\dots a_n}$} and \mbox{$\ket{b} = \ket{b_1\dots b_n}$}, where $\ket{a+b}$ denotes integer addition modulo $2^n$ \cite{cuccaro2004new,gidney2018halving}.

\begin{figure*}
    \centering
    \small
\includegraphics[width=\linewidth]{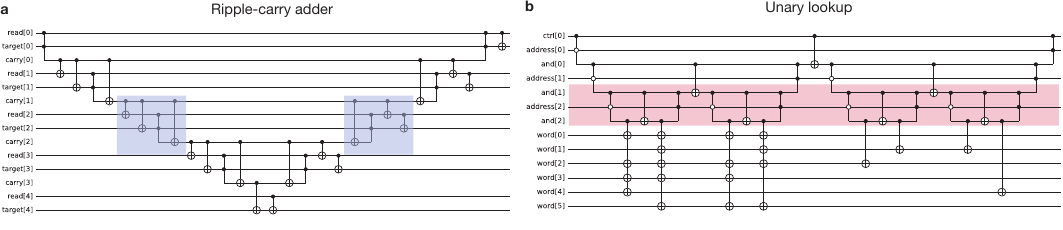}
    \caption{\textbf{Circuits for the ripple-carry adder and unary lookup.}
    \textbf{a,} We implement the ripple-carry adder using repeated small computation units (for example the blue boxes) so that each computation unit fits inside the processor codeblock.
    \textbf{b,} 
        The diagram shows an example lookup circuit on $3$ address qubits and $6$ word qubits. The words of the lookup table were generated randomly. Highlighted in red are three qubits: the least significant address qubit, and the two least significant ancilla AND qubits.
        At least half of the Toffolis and CNOTs acting on the address and ancilla AND qubits occur on these three registers, which means that at most half of the address and ancilla AND operations require I/O in our space efficient architecture.
    \label{fig:circuits}}
\end{figure*}

Because the space-efficient and balanced architectures use a small processor code block, they are best-suited for an adder circuit which operates only on a small number of logical qubits at a time. As a result, we analyze the Gidney variant~\cite{gidney2018halving} of the ripple-carry adder~\cite{vedral1996quantum, cuccaro2004new}. Shown in Extended Data Fig.~\ref{fig:circuits}a, the circuit acts on three qubit registers $a,b,c$, where $a$ are the offset qubits, $b$ are the target qubits, and $c$ are ancilla carry qubits. The circuit consists of two stages, which we refer to as the \emph{downwards pass} and the \emph{upwards pass} respectively. In the downwards pass, the circuit propagates the carries down from the least significant bit to the most significant bit, using Toffoli gates. In the upwards pass, the circuit uncomputes the carries using mid-circuit measurements, propagating from the most significant bit to the least significant bit. To perform $q_A$-bit addition, the circuit uses $3q_A$ qubits, $q_A$ Toffoli gates, and $q_A$ mid-circuit measurements.

Since in general the addition circuit will have too many logical qubits to fit inside the processor block ($\kproc < 3q_A$), we divide the adder into smaller sub-circuits as shown in Extended Data Fig.~\ref{fig:circuits}. We use $q_A / k_{\mathrm{add}}$ computation units for the downward and upward passes respectively, where we set $k_{\mathrm{add}} = \lfloor (\kproc - 1) / 3 \rfloor$. During each unit, we hold $k_{\mathrm{add}}$ input qubits, $k_{\mathrm{add}}$ output qubits, and $k_{\mathrm{add}} + 1$ carry qubits in the processor. For each computation unit of the downward pass, we only need to read $2k_{\mathrm{add}}$ qubits from the memory, while writing $3k_{\mathrm{add}}$ qubits back into memory, since the carry register is initialized to zero which can be done in the processor block. Similarly, for each computation unit of the upward pass, we read $3k_{\mathrm{add}}$ qubits but only need to write $2k_{\mathrm{add}}$ qubits back into memory, since the carry register is reset to zero. This gives an amortized input-output qubit count of $m_i = 2.5 k_{\mathrm{add}}$. For the downward pass, each unit involves $\beta_i = k_{\mathrm{add}}$ Toffolis and $\gamma_i = 0$ mid-circuit measurements, whilst for the upward pass $\beta_i = 0$ and $\gamma_i = k_{\mathrm{add}}$. This leads to an overall cost of
\begin{align}
    \tau_{\mathrm{adder}} &= \frac{q_A}{k_{\mathrm{add}}}\Big( (10 k_{\mathrm{add}} + 4k_{\mathrm{add}})
     +(10 k_{\mathrm{add}} + k_{\mathrm{add}})\Big)\taus \nonumber\\
    &= 25 q_A \taus.
\end{align}

Controlled adders require an additional $q_A$ Toffolis and $q_A$ mid-circuit measurements on the upwards pass \cite{gidney2018halving}, so an analogous calculation yields an overall cost of
\begin{align}
    \tau_{\mathrm{ctrl-adder}} &= \Big(\frac{q_A}{k_{\mathrm{add}}} (10 k_{\mathrm{add}} + 4k_{\mathrm{add}}) \nonumber\\
    &\quad + \frac{q_A}{k_{\mathrm{add}}} (10 k_{\mathrm{add}} + 4k_{\mathrm{add}} + 2k_{\mathrm{add}})\Big)\taus \nonumber\\
    &= 30 q_A \taus.
\end{align}

If $3q_A \leq \kproc$ then the entire adder fits inside the processor block. This is the case for the RSA circuit from \cite{gidney2025how} in the balanced architecture, since the maximum size of adder used is $33$ bits whilst the processor has $148$ qubits. In this case, we can simply teleport the entire $a$ and $b$ registers into the processor once at the beginning, and teleport them back once at the end. The total cost of the adder in this case is reduced to:
\begin{align}
    \tau_{\mathrm{adder}} = (8 q_A + 4 q_A + q_A)\taus = 13 q_A \taus.
\end{align}

\subsection{Time cost for lookup tables \label{app:lookups}}

Next we estimate the resource costs of performing lookups in the space-efficient and balanced architectures. A lookup has action
\begin{equation}
    |\ell\rangle |\psi\rangle \rightarrow |\ell\rangle \left(\bigotimes_{i=1}^{q_w} X_i^{T^{(\ell)}_i}\right) |\psi\rangle
\end{equation}
where $\{\vec{T}^{(\ell)}\}_{\ell=0}^{2^{q_a} - 1}$ is a classically-known lookup table with word length $q_w$ \cite{babbush2018encoding}.

We use the unary lookup circuits introduced in \cite{babbush2018encoding}. Shown in Extended Data Fig.~\ref{fig:circuits}, the circuit iterates through the address bitstrings $0,\dots,2^{q_a} - 1$ one-by-one. For each address bitstring, the circuit uses Toffolis to detect the identity of $\ell$, before applying the $X_i^{T^{(\ell)}_i}$ operators using CNOTs. The complete circuit uses $2q_a + q_w$ qubits, $2^{q_a}$ Toffolis, and $2^{q_a}$ mid-circuit measurements \cite{babbush2018encoding}.

If $2q_a + q_w \leq \kproc$, then the entire lookup fits inside the processor, and the cost is simply
\begin{align}
    \tau_{\mathrm{lookup}} &= (4(2q_a + q_w) + 4 \cdot 2^{q_a} + 2^{q_a}) \taus \nonumber\\
    &\approx (4 q_w / 2^{q_a} + 5) 2^{q_a} \taus \qquad\qquad (q_w \gg q_a).
\end{align}

For the balanced architecture, we assume that $2q_a < \kproc$ so that the processor can store the $2q_a$ qubits needed for the unary iteration with some space to spare to apply the word operators. In this case, we will split the lookup into $\big\lceil q_w/(\kproc - 2q_a) \big\rceil$ smaller lookups each involving all the address bits but a subset of at most $\kproc - 2q_a$ word bits. We then implement each of the smaller lookups using an I/O step, $2^{q_a}$ Toffolis, and $2^{q_a}$ mid-circuit measurements. We only need I/O between the memory and the processor after each small lookup to teleport the completed word qubits back into the memory, since the word qubits are always initialized to zero, which can be done in the processor without I/O. As such, the whole lookup takes time
\begin{align}
    \tau_{\mathrm{lookup}} &= \Big(\Big\lceil\frac{q_w}{\kproc - 2q_a}\Big\rceil \big(2(\kproc - 2q_a) + 4\cdot 2^{q_a} + 2^{q_a}\big)\Big)\taus \nonumber\\
    &\approx \frac{5 q_w}{\kproc - 2q_a} 2^{q_a} \taus \qquad\qquad\qquad (2^{q_a} \gg \kproc).
\end{align}
It should be noted that this compilation increases the total Toffoli count of the lookup by a factor of \mbox{$\big\lceil q_w/(\kproc - 2q_a) \big\rceil$}, since the Toffolis on the address qubits must be repeated for each of the $\big\lceil q_w/(\kproc - 2q_a) \big\rceil$ small lookups.

For the space-efficient architecture, the processor does not even have enough space to hold all of the address qubits, since $\kproc < 2q_a$. In this case, we designate 3 processor qubits to hold three address qubits that control the CNOTs acting on the word qubits, as shown in Extended Data Fig.~\ref{fig:circuits}b. The remaining $\kproc - 3$ processor qubits are used to store word qubits on which the CNOTs act. In this way, the lookup is split into $\big\lceil q_w / (k_p - 3) \big\rceil$ smaller lookups, each requiring $2^{q_a}$ Toffolis and $2^{q_a}$ mid-circuit measurements.

In this space-efficient compilation, many gates on the address qubits require I/O between processor and memory. In the complete unary iteration circuit, there are $2^{q_a}$ Toffolis, $2^{q_a}$ mid-circuit measurements, and $2^{q_a}$ CNOTs acting on the address qubits \cite{babbush2018encoding}. However, if we can hold 3 address qubits at any one time, then the structure of the unary iteration circuit implies that only half of the Toffolis, mid-circuit measurements, and CNOTs actually incur I/O cost---see Fig.~\ref{fig:circuits}. Moreover, each subsequent Toffoli or CNOT in the unary iteration circuit always overlaps with the previous one by at least one qubit, which means that the I/O cost is at most 2 qubits per Toffoli or mid-circuit measurement and at most 1 qubit per CNOT.

Overall, the cost of a lookup in the space-efficient architecture is:
\begin{align}
    \tau_{\mathrm{lookup}} &= \Big(\Big\lceil\frac{q_w}{\kproc - 1}\Big\rceil \big(2(\kproc - 1) + 4 \cdot (2 + 2 + 1)\cdot \frac{1}{2} \cdot 2^{q_a} \nonumber\\
    &\quad + 4\cdot 2^{q_a} + 2^{q_a}\big)\Big)\taus \nonumber\\
    &\approx \frac{15 q_w}{\kproc - 3} 2^{q_a} \taus \quad\qquad\qquad\qquad (2^{q_a} \gg \kproc).
\end{align}

\subsection{Time cost for \rsa\ and \ecc \label{app:time_cost_compilation}}

Shor's algorithm for \rsa\ and \ecc\ primarily uses adders and lookups as the dominant algorithmic subroutines \cite{jones2012layered, fowler2012surface, ogorman2017quantum, gheorghiu2019benchmarking, gidney2019how, dallairedemers2025brace, gidney2025how, zhou2025resource, webster2026pinnacle, chevignard2025reducing, proos2003shor, roetteler2017quantum, haner2020improved, litinski2023compute, gouzien2023performance, babbush2026quantum}. Based on the adder and lookup compilations developed in Sections \ref{app:adders} and \ref{app:lookups}, we now summarize and estimate the runtime of Shor's algorithm for \rsa\ and \ecc. We report the amortized time per Toffoli gate $\ttoff$ for each algorithm and each architecture in terms of the surgery time $\taus$; simply multiplying by the total Toffoli count will then yield the total runtime in terms of $\taus$.

\begin{table}[hbt!]
    \centering
    \resizebox{0.8\columnwidth}{!}{%
    \begin{tabular}{|c|cc|}
    \hline
     & \rsa & \ecc \\
    \hline
    Space-efficient & $\ttoff \approx 43 \taus$ & $\ttoff \approx 72 \taus$ \\
    Balanced & $\ttoff \approx 10 \taus$ & $\ttoff \approx 19 \taus$ \\
    \hline
    \end{tabular}}
    \caption{Amortized time-per-Toffoli for the \rsa\ and \ecc\ algorithms in the space-efficient and balanced architectures, all in terms of the time $\taus$ of a single surgery operation.}
    \label{tab:runtimes}
\end{table}

For the \rsa\ circuit in Ref.~\cite{gidney2025how}, approximately $50\%$ of the Toffolis arise from lookups on roughly 6 address bits with word-size at most $33$, and the remaining $50\%$ of Toffolis arise from adders on at most $33$ bits \cite{gidney2025how}. In the space-efficient architecture with $\kproc=10$, the time-per-Toffoli for adders and lookups is approximately $25\taus$ and $(15q_w/(\kproc - 3)) \taus \approx 71 \taus$ respectively. Overall this yields:
\begin{equation}
    \ttoff \approx 0.5 \cdot 25 \taus + 0.5 \cdot 71 \taus \approx 43 \taus
\end{equation}
for \rsa\ in the space-efficient architecture.

In the balanced architecture with $\kproc=148$, the small adders and lookups used in \rsa\ fit inside the processor, so the time-per-Toffoli for adders and lookups is only $13\taus$ and $(4 q_w / 2^{q_a} + 5) \taus \approx 7 \taus$ respectively. This yields:
\begin{equation}
    \ttoff \approx 0.5 \cdot 13 \taus + 0.5 \cdot 7 \taus \approx 10 \taus
\end{equation}
for \rsa\ in the balanced architecture.

For the \ecc\ algorithm, based on compilations in \cite{proos2003shor, roetteler2017quantum, haner2020improved, litinski2023compute, gouzien2023performance} we assume the following split in the Toffoli count between controlled-adders, adders and lookups:
\begin{itemize}
    \item $40\%$ $256$-bit adders,
    \item $50\%$ $256$-bit controlled-adders,
    \item $10\%$ lookups with $16$ address bits and word size $256$.
\end{itemize}
In the space-efficient architecture with $\kproc=10$, the time-per-Toffoli for adders and controlled-adders is $25\taus$ and $15\taus$ respectively, and the time-per-Toffoli for lookups becomes $(15q_w/(\kproc - 3)) \taus \approx 550 \taus$. Overall, for \ecc\ in the space-efficient architecture we get:
\begin{equation}
    \ttoff \approx 0.4 \cdot 25 \taus + 0.5 \cdot 15 \taus + 0.1 \cdot 550 \taus \approx 72 \taus
\end{equation}

In the balanced architecture with $\kproc=148$, the adders and controlled adders again have time-per-Toffoli $25\taus$ and $15\taus$ respectively, but now the time-per-Toffoli for lookups is $(5q_w/(\kproc - 2q_a)) \taus \approx 11 \taus$. Overall, for \ecc\ in the balanced architecture we get:
\begin{equation}
    \tau_{\mathrm{Tof}} \approx 0.4 \cdot 25 \taus + 0.5 \cdot 15 \taus + 0.1 \cdot 11 \taus \approx  19 \taus
\end{equation}

\section{Time-efficient architecture\label{app:time_efficient}}

Here we describe the time-efficient architecture and its associated resource costs for \rsa\ and \ecc. This architecture parallelizes Toffoli gates in core algorithmic subroutines, which can reduce time costs compared to the serial space-efficient and balanced architectures (Appendix~\ref{app:compilation}). As a proxy for the time reduction relative to these serial architectures, we consider replacing linear-depth ripple-carry addition~\cite{cuccaro2004new} used in prior resource estimate circuits~\cite{gidney2019how, gidney2025how, chevignard2025reducing, proos2003shor, roetteler2017quantum, haner2020improved, litinski2023compute, gouzien2023performance} with logarithmic-depth quantum carry-lookahead addition~\cite{draper2004logarithmic}. Similar linear-to-logarithmic-depth reductions are also possible in lookup tables~\cite{zhu2024unified}, which together with adders comprise the dominant Toffoli counts in prior circuits for Shor's algorithm~\cite{jones2012layered, fowler2012surface, ogorman2017quantum, gheorghiu2019benchmarking, gidney2019how, dallairedemers2025brace, gidney2025how, zhou2025resource, webster2026pinnacle, chevignard2025reducing, proos2003shor, roetteler2017quantum, haner2020improved, litinski2023compute, gouzien2023performance} (see Appendix~\ref{app:compilation}). As a result, we expect that the speedup between the ripple-carry and carry-lookahead adder is a reasonable proxy for the speedup of the full circuit.

To leverage these lower-depth primitives, we generate and consume magic $\ccz$ states in parallel using the high-rate $8T$-to-CCZ distillation protocol (Appendix~\ref{app:magic}), extended to larger factory codes encoding more logical qubits. In addition, we assume logical gadgets capable of measuring many logically disjoint PPMs on high-rate codes in parallel. Such capabilities have been demonstrated using, e.g. high-rate surgery techniques~\cite{zhang2025time, cowtan2025parallel, zheng2025high} (see also Appendix~\ref{app:surgery}), achieving parallel measurements of up to hundreds of logical operators on distance-$\sim$\,10 codes with ancilla overhead on the order of $1$–$2\times$ the code size~\cite{zhang2025time}. Ongoing work further develops these constructions for LP codes~\cite{zheng2026high}. We characterize variants of the time-efficient architecture by the \textit{parallelism} $P$, defined as the number of $\ccz$ states distilled and consumed simultaneously.

Our goal is not to specify concrete circuits or instruction sets, but to provide an approximate resource estimate for architectures supporting highly parallel Toffoli gates. The schemes outlined below rely on continued improvements in such parallel logical gadgets, which we expect to arise from further optimized high-rate surgery techniques~\cite{zheng2025high, cowtan2025fast, chang2026constant, zheng2026high} or alternative approaches such as homomorphic measurements~\cite{huang2022homomorphic, xu2025fast}. With this goal in mind, we first outline the architecture at a high level and then estimate its resource costs.

\subsection{Description}
In our time-efficient architecture, we perform logical operations directly on code blocks used for processing, thereby avoiding costly communication overhead with the memory blocks. Upcoming work finds codes at sufficiently high distance hosting $\gtrsim 100$ logical qubits with $20\%$ encoding rates, and $\gtrsim 600$ logical qubits with $30\%$ encoding rates~\cite{caltech2026codeatlas}, which we leverage for processing and as factory code blocks. We generate $P$ magic $\ccz$ states at a time using high-rate distillation with five factory code blocks and cultivated surface codes. As described later, we use different codes for processing and magic depending on the chosen value of $P$.

Using the carry-lookahead adder described in Ref.~\cite{draper2004logarithmic}, we consider both controlled and uncontrolled adders, which can each appear in cryptographic algorithms, depending on the particular circuit~\cite{gidney2019how, gidney2025how, chevignard2025reducing, proos2003shor, roetteler2017quantum, haner2020improved, litinski2023compute, gouzien2023performance}. In both cases, the circuits are comprised of $\approx$\,$4\log(n)$ parallel layers of Toffoli gates, each acting on a potentially different subset of logical qubits. This can be contrasted with ripple-carry addition, which requires $\approx1n$ and $\approx 2n$ Toffoli layers for uncontrolled and controlled addition respectively. The Clifford cost for the carry-lookahead adder is minimal---at most six layers of CNOTs in the entire adder---and therefore we neglect it in our estimates.

The carry-lookahead circuit~\cite{draper2004logarithmic} proceeds by first generating a group of $P$ magic states using the high-rate distillation described in Appendix~\ref{app:magic}. 
At large $P$, the space cost from surface codes can become large, so we allow the surface codes to be injected into the $\overline{T}$ block in a variable number of batches $n_\text{batch}$, resulting in a time cost per distillation of $8\times15n_\text{batch}$ (see Appendix~\ref{app:magic}). The resulting magic states are teleported into the computation using a single layer of parallel $\overline{ZZ}$ PPMs between the factory codes and the processor codes.
Subsequent $\overline{\text{CZ}}$ fix-ups on the processor codes are performed using two layers of parallel $\overline{ZZ}$ or $\overline{ZX}$ measurements with the assistance of up to $3P$ ancilla logical qubits~\cite{horsman2012surface}, which could come from e.g. the factory codes that are freed after the $\ccz$ teleportation.
Assuming each layer of parallel two-body PPMs can be implemented in parallel using, e.g. high-rate surgery, the total time cost of gate teleportation and fixups is therefore $3\taus = 2d$, where $d$ is taken to be 20 and $\taus$ denotes the surgery cycle time.
To reduce time costs, for layers involving fewer than five Toffoli gates, we delay the fixups until the next layer involving more than five Toffoli gates, at which point they are performed at negligible increase to the surgery system size.
At each step, the maximum possible number of Toffolis are implemented, corresponding to either $P$ (consuming a full factory) or the number of remaining Toffoli gates in the current Toffoli layer, whichever quantity is smaller. As soon as the produced $\ccz$ states are fully consumed, another batch is produced and the computation proceeds. 

\subsection{Resource costs\label{app:time_efficient_resource}}

We now estimate the space and time costs of this procedure for \ecc\ and \rsa\ at different levels of parallelism. The space cost is estimated from the sum of three quantities: the processor size, the resource zone size, and the operation zone size. As with the other architectures, our total qubit count includes data qubits and stabilizers in one basis ($X$ or $Z$); see Extended Data Table~\ref{tab:notation}. To estimate the processor size, we first compute the total number of logical qubits which need to be stored, given by the sum of the number of logical qubits from the original compilation~\cite{gidney2019how, babbush2026quantum} and the number of ancilla logical qubits required for carry-lookahead addition, $n-2\log(n)$, where $n$ is the number of bits in the adder~\cite{draper2004logarithmic}. We then divide by the encoding rate $r$ for the chosen $P$ to estimate the corresponding number of processor data qubits, from which we compute the total qubit count. For $P<600$ ($P\geq600$), we assume processor encoding rates of $r=20\%$ ($r=30$\%) based on upcoming work~\cite{caltech2026codeatlas}. Note that because we do not specify the concrete block size, these numbers are correct up to rounding errors of one block, which are small compared to the total qubit count.

For the resource zone, we require five factory blocks each encoding $P$ logical qubits which can be transversally coupled (note that each factory block can be assembled from smaller, independent code blocks). We assume these codes have the same rate as the processor, with the exception of $P<100$, where we assume a lower encoding rate of 4\% to allow for transversal coupling of smaller block sizes (e.g., the $\bb$ code). We also require $P/n_\text{batch}$ surface codes for $\lket{T}$ cultivation. Finally, we estimate the operation zone size by noting that at most $6P$ logical qubits are operated on in parallel at once (corresponding to three blocks for the $\ccz$ states, and 3$P$ processor logical qubits). 

We estimate the required ancilla size as $\gamma (6P/r)$, where $\gamma$ is the ratio between the ancilla size for measuring a layer of parallel PPMs and the size of the corresponding codes. For reference, \mbox{$\gamma \approx 1$–$2$ for \mbox{$d\sim 10$}} codes with $X$- or $Z$-type measurements using high-rate surgery~\cite{zheng2025high}. To account for larger codes and more complex PPMs, we consider \mbox{$\gamma = 1$–$3$} and plot results with \mbox{$\gamma = 2$} in Fig.~\ref{fig:resources}. We further assume that the ancilla cost for injecting $\lket{T}$ states during distillation does not exceed this bound. Because some physical qubits are freed up after the distillation factory terminates (including all of the surface code qubits and two factory blocks), these extra qubits are re-purposed for the operation zone for teleportations and fixups, further reducing its total size. We re-emphasize that the operation zone size is an estimate, and future work can benchmark and optimize logical gadgets for measuring parallel PPMs.

Now we estimate the space and time costs for a parallelism level $P$. For \rsa\ we consider \mbox{$P=100$} and \mbox{$P=1{,}160$} for which we choose \mbox{$n_\text{batch}=2$}. For \ecc\ we consider \mbox{$P=20$} and \mbox{$P=130$}, for which we choose \mbox{$n_\text{batch}=1$}. In the main text, we assume \mbox{$\gamma=2$}. We estimate the range in space costs due to surgery system size fluctuations by considering $\gamma=1$ and $\gamma=3$. This corresponds to approximate uncertainties of $67{,}000\substack{+4{,}000 \\ -3{,}000}$ qubits (\mbox{$P=100$}) and $102{,}000\substack{+28{,}000 \\ -4{,}000}$ qubits (\mbox{$P=1{,}160$}) for \rsa, and $19{,}000\substack{+3{,}000 \\ -2{,}000}$ (\mbox{$P=20$}) and $26{,}000\substack{+5{,}000 \\ 0}$ qubits (\mbox{$P=130$}) for \ecc. (The lower bound for the final estimate is zero because the space is dominated by the factory rather than code surgery.) We compute the factor speedup compared to the balanced architecture by comparing to the time cost of the ripple-carry adder, given by $\frac{2d}{3}\cdot25n$ and $\frac{2d}{3}\cdot2\cdot 15n$ for non-controlled and controlled addition, respectively. We take the ratio of these time estimates with the depth of the carry-lookahead adder, and take the average of the speedups for both controlled and non-controlled adders. The associated space and time costs are plotted in Fig.~\ref{fig:resources}b-c in the main text.

\vspace{1cm}
\noindent\textbf{Acknowledgments}
 M.E. and J.P. acknowledge support from the Institute for Quantum Information and Matter, an NSF Physics Frontiers Center (PHY-2317110).  M.E. acknowledges support from the NSF QLCI program (2016245). Q.X. acknowledges funding by the Walter Burke Institute for Theoretical Physics at Caltech.\\

\noindent\textbf{Competing interests}
The authors are shareholders of Oratomic, Inc., which is developing fault-tolerant quantum computers. M.C., Q.X, R.K., L.R.B.P, H-Y.H, and D.B. are full-time employees, and H.L., M.E., and J.P. are part-time employees, of Oratomic, Inc.
\\

\noindent\textbf{Correspondence and requests for materials} should be addressed to M.C., Q.X, and D.B.

\end{document}